\documentstyle[12pt,epsf,axodraw]{article}
\newcommand{\la}[1]{\label{#1}}
\newcommand{\be}{\begin{equation}}
\newcommand{\ee}{\end{equation}}
\newcommand{\ba}{\begin{eqnarray}}
\newcommand{\ea}{\end{eqnarray}}
\newcommand{\bi}{\begin{itemize}}
\newcommand{\ei}{\end{itemize}}
\newcommand{\rmi}[1]{{\mbox{\scriptsize #1}}}

\newcommand{\nr}[1]{(\ref{#1})}
\newcommand{\tr}{{\rm Tr\,}}
\newcommand{\hphi}{{\hat{\phi}}}
\newcommand{\nn}{\nonumber}
\newcommand{\ggg}{{g^2\over16\pi^2}}
\newcommand{\fr}[2]{{\frac{#1}{#2}}}
\newcommand{\msbar}{\overline{\mbox{\rm MS}}}
\newcommand{\lambdamsbar}{\Lambda_{\overline{\rm MS}}}
\newcommand{\dr}{{4d\to3d}}

\def\lsi{\raise0.3ex\hbox{$<$\kern-0.75em\raise-1.1ex\hbox{$\sim$}}}
\def\gsi{\raise0.3ex\hbox{$>$\kern-0.75em\raise-1.1ex\hbox{$\sim$}}}
\newcommand{\lsim}{\mathop{\lsi}}
\newcommand{\gsim}{\mathop{\gsi}}

\makeatletter \@addtoreset{equation}{section} \makeatother
\renewcommand{\theequation}{\arabic{section}.\arabic{equation}}

\begin{document}

\begin{titlepage}
\begin{flushright}
CERN-TH/97-81\\
BI-TP 97/10\\
HD-THEP-97-17\\
hep-ph/9704416\\
April 28, 1997\\
\end{flushright}
\begin{centering}
\vfill

{\bf 3D SU($N$) + ADJOINT HIGGS THEORY AND\\
FINITE TEMPERATURE QCD}
\vspace{0.8cm}

K. Kajantie$^{\rm a,b}$\footnote{keijo.kajantie@cern.ch},
M. Laine$^{\rm c}$\footnote{m.laine@thphys.uni-heidelberg.de},
K. Rummukainen$^{\rm d}$\footnote{kari@physik.uni-bielefeld.de} and
M. Shaposhnikov$^{\rm a}$\footnote{mshaposh@nxth04.cern.ch} \\

\vspace{0.3cm}
{\em $^{\rm a}$Theory Division, CERN, CH-1211 Geneva 23,
Switzerland\\}
\vspace{0.3cm}
{\em $^{\rm b}$Department of Physics,
P.O.Box 9, 00014 University of Helsinki, Finland\\}
\vspace{0.3cm}
{\em $^{\rm c}$Institut f\"ur Theoretische Physik,
Philosophenweg 16,\\
D-69120 Heidelberg, Germany\\}
\vspace{0.3cm}
{\em $^{\rm d}$Fakult\"at f\"ur Physik, Postfach 100131, D-33501
Bielefeld,
Germany}

\vspace{0.7cm}
{\bf Abstract}

\end{centering}

\vspace{0.3cm}\noindent
We study to what extent the three-dimensional SU($N$) + adjoint Higgs
theory can be used as an effective theory for finite temperature
SU($N$) gauge theory, with $N=2,3$. The parameters of the 3d theory are
computed in 2-loop perturbation theory in terms of
$T/\lambdamsbar,N,N_f$. The perturbative effective potential of the 3d
theory is computed to two loops for $N=2$. While the Z($N$) symmetry
probably driving the 4d confinement-deconfinement phase transition (for
$N_f=0$) is not explicit in the effective Lagrangian, it is partly
reinstated by radiative effects in the 3d theory. Lattice simulations
in the 3d theory are carried out for $N=2$, and the static screening
masses relevant for the high-temperature phase of the 4d theory are
measured. In particular, we measure non-perturbatively the 
$O(g^2 T)$ correction to the Debye screening mass. We find that
non-perturbative effects are much larger in the SU(2) + adjoint Higgs
theory than in the SU(2) + fundamental Higgs theory.
\vfill
\noindent

\end{titlepage}

\section{Introduction}

Corresponding to the known forces of nature there are two important
finite temperature phase transitions in elementary particle matter: the
QCD phase transition at $T=O(100)$ MeV and the EW (electroweak) phase
transition at $T=O(100)$ GeV. The former has been studied intensely
with numerical lattice Monte Carlo simulations~\cite{lat96} but, due to
the difficulties associated with treating dynamical quarks, conclusive
statements about the properties of the physical QCD transition cannot
yet be made. In contrast, for the EW case the problem can be
essentially solved \cite{isthere?} by a combination of analytic and
numerical means: first deriving by a perturbative computation
\cite{G}--\cite{mssm} a 3d effective theory $S_\rmi{eff}$ for the full
4d finite $T$ theory (``dimensional reduction'') and then solving this
confining non-perturbative 3d theory by numerical means
\cite{fkrs}--\cite{mt}. The effective field theory approach has been
intensively used for computations in high temperature QCD
\cite{huanglissia}--\cite{BN2}, as well.

In fact, there has been no doubt that dimensional reduction of QCD
would work well at very high temperatures in the QCD plasma phase,
$T\gg T_c$, in the sense of the 3d theory giving correctly the static
correlation functions of the theory. The situation is quite different
in the phase transition region $T\approx T_c$: the effective finite
temperature gauge coupling $g^2(T)$ is becoming large so that the mass
hierarchy $g^2 T, gT \ll \pi T$ needed for the construction of a simple
local effective theory is lost. In particular, quarks play dynamically
a crucial role and it seems that an effective theory using constant (in
imaginary time) field configurations as essential degrees of freedom
cannot be accurate (quarks are antiperiodic and are thus integrated out
in this effective theory, their effect appearing only in the 3d
parameters). However, it still seems well motivated to study how far
one can actually go with dimensionally reduced QCD towards the phase
transition region and the purpose of this paper is to do this. In
comparison with earlier work~\cite{nadkarni3d}--\cite{reisz2} we go
further firstly by determining the two continuum parameters of the
effective theory in terms of $T/\Lambda_\rmi{QCD},N,N_f$ (quark masses
are neglected) using 2-loop perturbation theory and the
techniques developed in~\cite{generic}. Secondly,
we are now able to extrapolate the lattice results to the continuum
limit, using the lattice--continuum relations derived in
\cite{fkrs,mlaine2} (see also~\cite{moore_a}). Our conclusions will
thus be different from those of the previous lattice
studies~\cite{nadkarni3d}--\cite{reisz3} for SU(2).

Quite independent of its use as a finite $T$ effective theory, the 3d
SU(2)+adjoint Higgs theory is interesting because it has monopoles
\cite{thooftmon,polyakovmon} which remove the photon from the physical
spectrum and replace it by a pseudoscalar with the mass
\cite{polyakovphoton}
\be
{m_\gamma^2\over g_3^4}\sim \exp\biggl[-{4\pi m_W\over g_3^2}\biggr]
\sim \exp\biggl[-{M_\rmi{mon}\over T}\biggr],
\la{gammamass}
\ee
where $m_W$ is the perturbative mass of the vector excitation in the
broken phase (in a gauge invariant analysis $m_W$ does not correspond
to a physical state). A measurement of the photon mass with gauge
invariant operators will permit one to make statements about the monopole
density in the broken and symmetric phases. These questions have
recently been addressed in~\cite{ph}.

The paper is organized as follows. In Sec.~2 we discuss the derivation
of the effective 3d theory. In Sec.~3 we perform some perturbative
estimates within that theory. In Sec.~4 we address the
role of the Z($N$) symmetry in the effective 3d theory, and in Sec.~5
we consider the 3d lattice results. The implications of the analysis
for the 4d finite temperature gauge theory are in Sec.~6, and the
conclusions in Sec.~7.

\section{Relating the 4d and 3d theories}

Finite $T$ QCD (with $N$ colours for the moment) is defined by the
action
\be
S=\int_0^\beta\!d\tau\int\! d^3x\biggl\{\fr14 F_{\mu\nu}^aF_{\mu\nu}^a+
\sum_i\bar\psi_i[\gamma_\mu D_\mu(A)+m_i]\psi_i\biggr\},
\ee
where
\be
F_{\mu\nu}^a=\partial_\mu A_\nu^a-\partial_\nu A_\mu^a-
gf_{abc}A_\mu^bA_\nu^c,
\ee
\be
D_\mu=\partial_\mu+igA_\mu,\quad A_\mu=A_\mu^aT^a,
\ee
and where the gluon field is periodic (quark field antiperiodic) in
$\tau$ with period $\beta=1/T$. In the matrix representation
\be
F_{\mu\nu}=\partial_\mu A_\nu-\partial_\nu A_\mu+ig[A_\mu,A_\nu]=
(ig)^{-1}[D_\mu,D_\nu].
\ee
After renormalisation (we use the $\msbar$ scheme), $g^2$ becomes
scale dependent, and to 1-loop
\be
g^{-2}(\mu)={11N-2N_f\over 24\pi^2}\ln{\mu\over
\lambdamsbar}.
\la{running}
\ee
Some useful group theoretical relations for the SU(N) generators
are given in Appendix~A.

Upon dimensional reduction, the field $A_0^a$ becomes an
adjoint Higgs field and the general form of the super-renormalizable
Lagrangian of the
3d effective theory can be written down:
\ba
L_\rmi{eff}[A_i^a,A_0^a] & = &  \fr14  F_{ij}^aF_{ij}^a
+ \tr [D_i,A_0][D_i,A_0]  \nn \\
& + & m_D^2\tr A_0^2 +\lambda_A(\tr A_0^2)^2+
\bar\lambda_A[\tr A_0^4-\fr12(\tr A_0^2)^2].
\la{leff}
\ea
The reduction process does not generate terms proportional to
$\tr A_0^3$. Operators of higher dimensions
are parametrically of the form $g^6A_0^6$ and
can be neglected relative to the retained term $g^4A_0^4$
as long as $A_0\ll 2\pi T/g$ (their
contributions to the correlators we will measure
are also of higher order).

Note that at very high temperatures, it is even possible
to integrate out the
$A_0$-field~\cite{appelqpisarski,perturbative,generic,BN2}.
Here we want to go as close to $T_c$ as possible,
and hence we keep $A_0$ in the effective Lagrangian.

For general $N$ two independent quartic couplings appear in
eq.~\nr{leff}.
We shall take in this paper $N=2,3$ for which eq.~\nr{su2su3} implies
that one can take $\bar\lambda_A=0$. The case $N=5$ with both couplings
is treated in \cite{rajantie5}.

The effective theory thus depends on three
dimensionful couplings: $g_3^2$, $m_D^2$, $\lambda_A$. Instead of
these one can use one dimensionful scale and two
dimensionless parameters, chosen as
\be
g_3^2,\quad y={m_D^2(g_3^2)\over g_3^4},\quad
x={\lambda_A\over g_3^2}.
\la{variables}
\ee
The 3d theory is super-renormalizable and
$g_3^2,\lambda_A$ are renormalisation scale independent
while $m_D^2$ is of the form
\ba
m_D^2(\mu)&=&{f_{2D}\over16\pi^2}\ln{\Lambda_D\over\mu},\nn\\
f_{2D}&=&2(N^2+1)(Ng_3^2-\lambda_A)\lambda_A,
\la{md}
\ea
where $\Lambda_D$ is a constant specifying the theory.
The scale in the definition of $y$ in eq.~\nr{variables} is chosen
as the natural one, $g_3^2$. It is noteworthy that there is no
$g_3^4$-term in \nr{md}.

The process of dimensional reduction now implies finding the
relation between the physical parameters of finite temperature QCD and
$g_3^2,y,x$. The Lagrangian parameters of QCD are after renormalisation
$\lambdamsbar,m_i(\mu)$; the physical parameters are hadron
masses (one mass to set the scale
and the rest as dimensionless mass ratios).
As hadron masses are entirely non-perturbative, it is conventional to
use $\alpha_S(m_Z)$ and define $\lambdamsbar$ as the scale for
which $\alpha_S$ diverges when evolved to smaller scales.

We will derive the parameters of the 3d theory so that
the relative errors are of the order $O(g^4)$. This requires
a 1-loop calculation for the gauge coupling, but for
the mass parameter $m_D^2$ and the scalar coupling
$\lambda_A$ one needs a 2-loop derivation. The actual reliability
of this calculation is to be discussed below.

To 1-loop (tree level for $g_3^2$)
the calculation gives \cite{nadkarnidr,land}
\ba
g_3^2&=&g^2(\mu)T, \\
m_D^2&=&\fr13\Bigl(N+\fr12N_f\Bigr)g^2(\mu)T^2,\\
\lambda_A&=&(6+N-N_f){g^4(\mu)T\over24\pi^2},\\
\bar\lambda_A&=&(N-N_f){g^4(\mu)T\over12\pi^2}.
\ea
We take $N_f$ flavours of massless quarks, although the
dependence on mass thresholds could also,
in principle, be included.

At this level the scale $\mu$ is unspecified; thus a 2-loop
derivation of the effective theory, constituting a resummation
of the perturbative series and establishing the scale at which
the 3d parameters are to be evaluated, is needed.
Indeed, the 3d couplings are
scale independent (note that the 3d
scale dependence in eq.~\nr{md} is of order $g^6$ and thus does
not yet enter at this level of the 4d$\to$3d reduction) so that
the $\mu$ dependence of the 2-loop result must be the
following (we only discuss $N=2,3$ so that
$\bar\lambda_A$ is irrelevant):
\ba
g_3^2&=&g^2(\mu)T\Bigl[1+\ggg(L+c_g)\Bigr], \la{g3g}\\
m_D^2&=&\fr13\Bigl(N+\fr12N_f\Bigr)
g^2(\mu)T^2\Bigl[1+\ggg(L+c_m)\Bigr],\la{mdg}\\
\lambda_A&=&(6+N-N_f){g^4(\mu)T\over24\pi^2}
\Bigl[1+2\ggg(L+c_l^{(N)})\Bigr],
\la{lag}
\ea
where
\be
L=\fr{22N}3\ln{\mu\over\mu_T}-{4N_f\over3}\ln{4\mu\over\mu_T},
\ee
and the $c_i$ are constants to be found.

The derivation of the parameters can be most easily made using
the background field gauge. Calculating to 1-loop the
contribution from the $n\neq 0$ modes to the zero mode
correlators, one gets the relations between the 3d and 4d fields:
\ba
(A_0^aA_0^b)^{\rm 3d} \!\!\! & = & \!\!\! \frac{1}{T}
(A_0^aA_0^b)(\mu)\biggl\{
1+\frac{g^2}{16\pi^2}\biggl[
-\frac{N}{3}\biggl(22\ln\frac{\mu}{\mu_T}+6\xi-7
\biggr)
+\frac{N_f}{3}\biggl(
4\ln\frac{4\mu}{\mu_T}-2
\biggr)
\biggr]
\biggr\}, \nn \\
(A_i^aA_j^b)^{\rm 3d} \!\!\! & = & \!\!\! \frac{1}{T}
(A_i^aA_j^b)(\mu)\biggl\{
1+\frac{g^2}{16\pi^2}\biggl[
-\frac{N}{3}\biggl(22\ln\frac{\mu}{\mu_T}+1
\biggr)
+\frac{N_f}{3}\biggl(
4\ln\frac{4\mu}{\mu_T}\biggr)
\biggr]
\biggr\}, \la{fields}
\ea
where $\xi$ is the gauge parameter and
\be
\mu_T=4\pi e^{-\gamma_E}T\approx 7.0555T
\ee
is the standard thermal scale arising in perturbative reduction in the
$\msbar$ scheme~\cite{perturbative,huanglissia}. The 3d gauge coupling
can be read directly from the gauge independent $A^a_i A^b_j$
normalization factor in this gauge~\cite{huanglissia}. For the other
parameters, one needs the 2-loop correlators at zero momenta of the
$A_0$-fields in the 4d and 3d theories, to separate the contributions
coming from the $n\neq 0$ modes. These can be most easily derived with
the effective potential. The full 4d 1-loop effective potential can be
read from eqs.~(4.2) (with $3/4\to 4/3$), (5.2) of~\cite{cka}, and the
2-loop effective potential from eqs.~(4.3), (5.4) of~\cite{cka}. The
1-loop potential is gauge independent, but the 2-loop potential in
these equations corresponds to $\xi=3$, since the extra rescalings
added in eqs.~(3.17), (3.18) of~\cite{cka} vanish for that gauge
parameter. For the mass parameter $m_D^2$ one has to subtract the
contribution from 3d, which has to be calculated separately, but for
$\lambda_A$ there is no 3d contribution in the 4d 2-loop effective
potential and thus the coefficient of the quartic term gives directly
the $n\neq 0$ contribution. Finally, to get the 2-loop results for the
3d parameters, one needs the terms arising when the rescalings
in~\nr{fields} combine with the 1-loop results. Alternatively, the
2-loop mass parameter $m_D^2$ can be read directly from~\cite{BN2}.
Note that the 2-loop cubic terms in the 4d effective potential are not
used in the derivation of the 3d parameters and are not reproduced by
the 3d theory (they are higher order contributions when $\delta A \lsim
\pi T$).

The computation described above gives
\ba
c_g&=&\fr{N}3,\la{cg}\\
c_m&=&{10N^2+2N_f^2+9N_f/N\over 6N+3N_f},\\
c_l^{(2)}&=&{7/3-109N_f/96\over 1-N_f/8}+\fr23 N_f,\\
c_l^{(3)}&=&{7/2-23N_f/18\over 1-N_f/9}+\fr23 N_f.\la{cl3}
\ea
Note that for the computation of the free energy of QCD in the
symmetric phase to order $g^5T^4$ through a 3d theory~\cite{BN2}, one
only needs $c_m$; the constants $c_g$ and $c_l^{(N)}$ give higher order
contributions.

Another useful representation of eqs. (\ref{g3g})--(\ref{lag}) can be
derived by choosing different renormalization scales for different
parameters in a way that 1-loop corrections to the 3d parameters
vanish. In general, the solution of $L+c=0$ is
\be
\mu=\mu_T\exp\biggl({-3c+4N_f\ln4\over22N-4N_f}
\biggr)\equiv \mu_T\hat\mu. \la{muopt}
\ee
Using \nr{running} and \nr{cg} this gives (for $N_f=0$):
\ba
{g_3^2\over T}&=&
{24\pi^2\over11N\ln(6.742T/\lambdamsbar)}.
\la{g3f0}
\ea
Secondly, for $x$ one obtains
\be
x={\lambda_A\over g_3^2}={6+N-N_f\over11N-2N_f}\,\,{1\over
\ln(\mu_T\hat\mu/\lambdamsbar)},
\ee
where $\hat\mu\,\,(=e^{-3/11}$ for $N_f=0$) is obtained from
\nr{muopt} with $c=2c_l^{(N)}-c_g$. This equation
gives quantitatively the relation between the 4d theory variables
$T/\lambdamsbar,N,N_f$ and the 3d effective theory variable $x$.
For $N_f=0,\,\,N=2,3$ one has
\be
x={6+N\over11N}\,\,{1\over\ln(5.371T/\lambdamsbar)}.
\la{xT}
\ee

Finally, concerning $y=m_D^2(g_3^2)/g_3^4$ note that the
RHS of eqs.~\nr{g3g}--\nr{lag} only depends
on $g^2(\mu)$ and $N_f$. Eliminating $g^2(\mu)$ one obtains
a $\mu$-independent
relation between the dimensionless scale independent variables
$y,x$. To leading order $y=m_D^2/g^4_3
\sim 1/g^2; x=\lambda_A/g_3^2\sim g^2$ so that $y\sim 1/x$. The
complete relation is
\ba
y_\dr^{(N=2)}(x)&=&{(8-N_f)(4+N_f)\over144\pi^2x}+
{192-2N_f-7N_f^2-2N_f^3\over96(8-N_f)\pi^2}+{\cal O}(x),\la{y_dr2}\\
y_\dr^{(N=3)}(x)&=&{(9-N_f)(6+N_f)\over144\pi^2x}+
{486-33N_f-11N_f^2-2N_f^3\over96(9-N_f)\pi^2}+{\cal O}(x).\la{y_dr3}
\ea
For $N_f=0$ these have the simple forms
\ba
y_\dr^{(N=2)}(x)&=&{2\over9\pi^2x}+{1\over4\pi^2}={2\over9\pi^2x}
\biggl(1+\fr98x+{\cal O}(x^2)\biggr),\la{ydrntwo}\\
y_\dr^{(N=3)}(x)&=&{3\over8\pi^2x}+{9\over16\pi^2}={3\over8\pi^2x}
\biggl(1+\fr32x+{\cal{O}}(x^2)\biggr).
\ea

With the use of known lattice data for the phase transition in
pure gauge theories in 4d for $N=2,3$ one can define the value of $x$
corresponding to the critical temperature. We have: $T_c/\lambdamsbar =
1.23(11)$ for SU(2) and $T_c/\lambdamsbar = 1.03(19)$ for SU(3)
according to~\cite{fhk}. Thus, for $N=2$ the value of $x$ corresponding
to $T_c$ is about $x_c=0.2$ ($x_c \simeq 0.17$ for SU(2) and $x_c \simeq
0.14$ for SU(3)). To have a feeling of the accuracy required for the
assessment of the lattice results below, the leading value of $y(0.2)$
is 0.1126 to which a 22.5\% 2-loop correction 0.0253 is to be added.
One might then estimate that the next (omitted) term is roughly 5\% so
that the theoretical result is $y(x=0.2)=0.138(6)$.

Quite surprisingly, the power series defining the mapping of the 4d
parameters to the 3d ones seems thus to be quite convergent even at the
critical temperature. However, this does not prove that the 3d theory
can adequately describe the confinement-deconfinement phase transition
at high temperatures. There is another important criterion for the
applicability of dimensional reduction, namely that the typical 3d
mass scale must be much smaller than $\pi T$, since only then the
integration out of the non-zero Matsubara modes is self-consistent.
As we will see, it is this point which does not allow an effective
3d description near the critical point.

Summarizing, to the extent that finite $T$ QCD can be regarded as an SU(N)
gauge theory with $N_f$ massless quarks characterized by the physical
quantities $T/\lambdamsbar,N,N_f$, a 3d effective theory given by the
super-renormalizable Lagrangian \nr{leff} with the couplings
$g_3^2,y,x$ (eqs.~\nr{variables}) can be derived. The relation between
these two sets is in eqs.~\nr{g3f0}, \nr{xT}, \nr{y_dr2} and
\nr{y_dr3}.

Note the crucial difference in comparison with SU(N) + fundamental
Higgs theories, in which there is a Higgs potential with two parameters
already at the tree level. Then (for $N=2$) the variable $x$ (the Higgs
self-coupling in the effective theory) is essentially 
the zero temperature Higgs mass
and $y$ (the scalar mass in the effective theory) 
is $\sim (T-T_c)/T_c$. The whole $x>0$ plane
corresponds to some physical SU(2)+Higgs finite $T$ theories. For pure
SU(2), in contrast, both $x$ and $y$ are given by
$T/\lambdamsbar,N,N_f$ and for fixed $N,N_f$ only one curve $y_\dr(x)$
corresponds to a physical 4d theory. Assume the effective theory has a
phase transition along some curve $y=y_c(x)$, which we shall soon
determine with lattice Monte Carlo simulations. Then this transition
could correspond to a physical 4d transition only if it intersects the
curve $y_\dr(x)$ in a region where the derivation of the effective
theory is reliable.

\section{Perturbation theory in 3d for $N=2$}

The aim now is to study the phase structure of the 3d theory defined by
the Lagrangian \nr{leff} on the plane of its variables $x,y$. On the
full quantum level the study has to be based on gauge invariant
operators and will be carried out by numerical means in Sec.~5; only
this gives reliable answers. However, it is also quite useful to fix
the gauge and study the problem perturbatively. On the tree level the
answer then is obvious: there is a symmetric phase for $y>0$
($m_D^2>0$) at all $x>0$ and a broken phase for $y<0$. To get a more
accurate result, one shifts the field by
\be
A_0^a\to A_0^a+\phi\delta_{a3},
\ee
obtains 3=1+2 scalars with masses
\ba
m_1^2&=&m_D^2+3\lambda_A\phi^2=g_3^4(y+3x\hphi^2),\nonumber \\
m_2^2&=&m_D^2+\lambda_A\phi^2=g_3^4(y+x\hphi^2),
\la{scalarm}
\ea
two massive vectors with mass
\be
m_T^2=g_3^2\phi^2=g_3^4\hphi^2,
\la{vectorm}
\ee
and one massless vector. The 1-loop potential in the Landau
gauge is 
\ba
V_{\rm 1-loop} \!\!\! & = & \!\!\!
\fr12m_D^2(\mu)\phi^2+\fr14\lambda_A\phi^4-{1\over12\pi}
\Bigl[4m_T^3+m_1^3+2m_2^3\Bigr]\la{v1loop}\\
& = & \!\!\!
g_3^6\biggl\{\fr12y(\mu)\hphi^2+\fr14x\hphi^4-{1\over12\pi}
\Bigl[4\hphi^3+(y+3x\hphi^2)^{3/2}+2(y+x\hphi^2)^{3/2}\Bigr]
\biggr\},\nn
\ea
where
\be
y(\mu)=y+{1\over16\pi^2}(20x-10x^2)\ln{g_3^2\over\mu}.
\ee

Let us consider first the high temperature limit of the 1-loop
effective potential. It corresponds to the case $x\to0$. The leading
terms are:
\be
V_{\rm 1-loop}/g_3^6=\fr14x\hphi^2\biggl[\biggl(\hphi-{2\over3\pi x}
\biggr)^2+{2y\over x}\biggl(1-{2\over9\pi^2xy}\biggr)\biggr].
\la{vonlyvector}
\ee
Two degenerate states are obtained when the last term vanishes.
{}From this one finds that
\ba
y_c(x)&=&{2\over9\pi^2x},\la{yc2}\\
\hphi_\rmi{symm}&=&0,\quad\hphi_\rmi{broken}={2\over3\pi x},\\
\sigma_3&=&{2^{3/2}\over81\pi^3x^{5/3}},
\ea
where $\sigma_3$ is the 3d interface tension divided by $g_3^4$.
Further, the upper and lower metastability points $y_\pm(x)$
are given by
\be
y_+(x)={1\over4\pi^2x},\quad y_-(x)=0.
\ee

The contributions to the 2-loop potential
from vectors (%
\begin{minipage}{0.55cm}
\begin{picture}(15,10)(0,0)
\Photon(0,5)(15,5){1.5}{2.5}
\end{picture}
\end{minipage}%
),
ghosts (%
\begin{minipage}{0.55cm}
\begin{picture}(15,10)(0,0)
\Line(0,4)(15,4)
\Line(0,6)(15,6)
\end{picture}
\end{minipage}%
) and
scalars (%
\begin{minipage}{0.55cm}
\begin{picture}(15,10)(0,0)
\Line(0,5)(15,5)
\end{picture}
\end{minipage}%
) are (without a factor $1/(16\pi^2)$)
\ba
\begin{minipage}{1.2cm}
\begin{picture}(30,30)(0,0)
\Photon(0,15)(30,15){1.5}{4}
\PhotonArc(15,15)(15,0,360){1.5}{12}
\end{picture}
\end{minipage}
&=&g_3^2m_T^2[4H(m_T,m_T,0)+H(m_T,0,0)-{89\over24}], \la{vvv} \\
\begin{minipage}{1.2cm}
\begin{picture}(30,30)(0,0)
\Photon(0,15)(30,15){1.5}{4}
\CArc(15,15)(13,0,360)
\CArc(15,15)(15,0,360)
\end{picture}
\end{minipage}
&=&\fr12g_3^2m_T^2H(m_T,0,0),\\
\begin{minipage}{1.6cm}
\begin{picture}(40,20)(0,0)
\PhotonArc(10,10)(10,0,360){1.2}{9}
\PhotonArc(30,10)(10,-180,180){1.2}{9}
\end{picture}
\end{minipage}
&=&\fr43g_3^2m_T^2,\\
\begin{minipage}{1.2cm}
\begin{picture}(30,30)(0,0)
\Photon(0,15)(30,15){1.5}{4}
\CArc(15,15)(15,0,360)
\end{picture}
\end{minipage}
&=&-\fr12g_3^2[D_{SSV}(m_2,m_2,0)+2D_{SSV}(m_1,m_2,m_T)],\\
\begin{minipage}{1.6cm}
\begin{picture}(40,20)(0,0)
\CArc(10,10)(10,0,360)
\PhotonArc(30,10)(10,-180,180){1.2}{9}
\end{picture}
\end{minipage}
&=&2g_3^2m_T(m_1+m_2), \\
\begin{minipage}{1.2cm}
\begin{picture}(30,30)(0,0)
\Line(0,15)(30,15)
\PhotonArc(15,15)(15,0,360){1.5}{12}
\end{picture}
\end{minipage}
&=&-g_3^4\phi^2[D_{VVS}(0,m_T,m_2)+2D_{VVS}(m_T,m_T,m_1)], \\
\begin{minipage}{1.2cm}
\begin{picture}(30,30)(0,0)
\Line(0,15)(30,15)
\CArc(15,15)(15,0,360)
\end{picture}
\end{minipage}
&=&-\lambda_A^2\phi^2[3H(m_1,m_1,m_1)+2H(m_1,m_2,m_2)],\\
\begin{minipage}{1.6cm}
\begin{picture}(40,20)(0,0)
\CArc(10,10)(10,0,360)
\CArc(30,10)(10,-180,180)
\end{picture}
\end{minipage}
&=&\fr14\lambda_A[3m_1^2+4m_1m_2+8m_2^2], \la{vvs}
\ea
where
\ba
&& \hspace*{-0.5cm} D_{SSV}(m_1,m_2,M)=\nn\\
&&(M^2-2m_1^2-2m_2^2)H(M,m_1,m_2)+
{1\over M}(m_1+m_2)[M^2+(m_1-m_2)^2]-m_1m_2\nn\\
&&+{1\over M^2}
(m_1^2-m_2^2)^2[H(M,m_1,m_2)-H(0,m_1,m_2)],
\ea
\ba
&& \hspace*{-0.5cm} D_{VVS}(M_1,M_2,m_3)=\nn\\
&&\fr32H(M_1,M_2,m_3)+{1\over4M_1M_2}
[(M_1-M_2)^2-m_3(M_1+M_2+m_3)]\nn\\
&&+{1\over4M_1^2M_2^2}\biggl\{m_3^4[H(M_1,M_2,m_3)-H(M_1,m_3,0)-
H(M_2,m_3,0)+H(m_3,0,0)]\nn\\
&&+(M_1^4-2M_1^2m_3^2)[H(M_1,M_2,m_3)-H(M_1,m_3,0)]\nn\\
&&+(M_2^4-2M_2^2m_3^2)[H(M_1,M_2,m_3)-H(M_2,m_3,0)]\biggr\}.\la{dvvs}
\ea
The sunset function is
\be
H(m_1,m_2,m_3)={1\over16\pi^2}
\biggl(\ln{\mu\over m_1+m_2+m_3}+\fr12\biggr).
\ee
In \nr{vvv}--\nr{dvvs} the factor $1/(16\pi^2)$
appearing in $H$  should be suppressed.

The use of 2-loop effective potential allows the construction of the
critical curve $y_c(x)$ beyond 1-loop level. The  numerical
computation is shown in Fig.~\ref{ycritical}.

\begin{figure}[t]

\vspace*{-2.0cm}

\hspace{1cm}
\epsfysize=18cm
\centerline{\epsffile{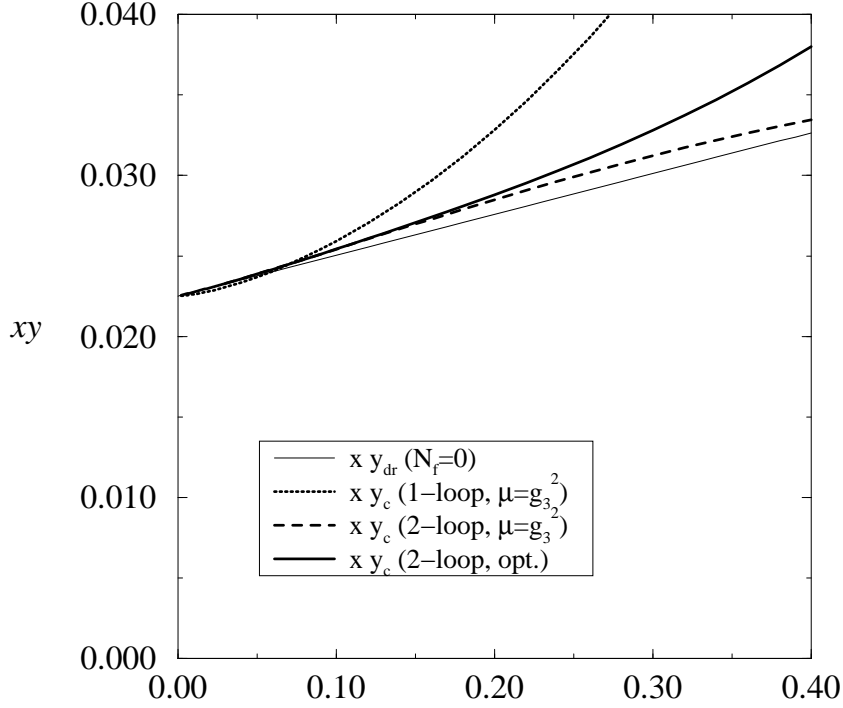}}

\vspace*{-7cm}

\caption[a]{The critical curve $y=y_c(x)$ (multiplied by $x$)
computed from 1- and 2-loop
potentials in the SU(2)+adjoint Higgs theory. Two of the curves are
plotted
for the scale choice $\mu=g_3^2$; $\mu$ for the second 2-loop curve is
determined by optimization~\cite{perturbative}.
The scale dependence should give an estimate of
higher order corrections. The thin  curve is $y_\dr(x)$
from eq.~\nr{ydrntwo}. For $x\to0$ all curves approach $2/(9\pi^2)=
0.0225$. The non-perturbative critical curve computed numerically
is given in Fig.~\ref{ycdata}.}
\la{ycritical}
\end{figure}

\section{Z($N$) symmetry in the 3d theory}

For $N_f=0$, hot SU($N$) gauge theory exhibits a Z($N$) symmetry
\cite{gpy,cka} so that there are $N$ equivalent ground states. In one
of them $A_0=0$, in the others $A_0^{N^2-1}\sim 2\pi T/g$. In the weak
coupling limit the field in the other minima thus becomes large. One
can associate the QCD phase transition with the breaking of this
symmetry so that at high temperatures one sits in one of the equivalent
minima, say $A_0=0$. As one goes to lower temperatures, the barrier
between the minima is becoming lower and at $T_c$, the symmetry is
supposed to get restored. If one wants to apply a 3d theory down to
temperatures close to $T_c$, it is thus important to discuss the role
of the Z($N$) symmetry in~3d.

The construction of the super-renormalizable effective 3d field theory
in Sec.~2
requires small amplitudes of the adjoint field $A_0$, $A_0 \ll 2\pi
T/g$. In practice, this means that one works around the $A_0=0$
minimum in the reduction step, so that the effective theory describes
reliably only fluctuations of magnitude $\delta A \lsim \pi T$ around
that minimum. Thus, the price one pays for the simple effective theory
is that the Z($N$) symmetry is not reproduced by it.

Some remnant of the Z($N$) symmetry nonetheless remains in the 3d
theory. The effective theory will on its critical curve still contain
$N$ metastable states but these are not completely equivalent, e.g.,
the correlators measured in them are different. As discussed,
only the correlators in the symmetric phase ($A_0=0$) reliably
represent 4d physics.

Indeed, within the interval $0<A_0^3<2\pi T/g$, eq.~\nr{vonlyvector} at
$y_c$ is seen to be the same as the 4d 1-loop potential in an $A_0^3$
background:
\ba
V(A_0^3)&=&{2\pi^2\over3}T^4\biggl[B_4(0)+2B_4
\biggl({gA_0^3\over2\pi T}\biggr)\biggr]
\la{v4d}\\
&=&-{6\pi^2\over90}T^4+\fr13g^2T^2(A_0^3)^2
+{1\over12\pi^2}g^4(A_0^3)^4-{1\over3\pi}g^3T(A_0^3)^3
,\quad0<A_0^3<{2\pi T\over g},\nn
\ea
where
\be
B_4(x)\equiv-{1\over30} + (x\,{\rm mod}1)^2[1-(x\,{\rm mod}1)]^2.
\ee
The effective potential \nr{v4d} exhibits the Z(2) symmetry of the full
4d theory as the periodicity in $A_0^3$ with period $2\pi T/g$; this
symmetry is not explicit in the effective theory on the tree level, but
it is partly reinstated by radiative effects 
as seen in eq.~\nr{vonlyvector}.
This is not surprising: in
deriving the 4d 1-loop potential one performs a frequency sum over
$n=0,\pm1,\pm2,...$. In deriving the couplings of the 3d theory one
performs basically a frequency sum over $n=\pm1,\pm2,...$, and the
$n=0$ mode enters when computing the effective potential within the 3d
theory. Note also that~\nr{yc2} is exactly the same as the leading term
in \nr{ydrntwo}.

Thus, while the Z(2) symmetry is not explicit in the effective action,
the second degenerate minimum is generated radiatively. Even with all
higher order corrections in the non-perturbative solution of the 3d
theory there will be two (for $N=2$) degenerate minima at some values
of the 3d parameters, but this will no longer take place at any
temperature $T>T_c$. Moreover, while these states are completely
equivalent in the original 4d theory, the correlators measured in them
will be different in the effective theory, the symmetric phase being
the physical one. For SU(3) the full 4d theory has 3 equivalent states
while in the 3d theory presumably only two of them have the same
correlators; these are the unphysical ones corresponding to a large
value $\sim 2\pi T/g$ of the fields. 

\section{Lattice analysis}

\subsection{Discretization}

The lattice action corresponding to the continuum theory \nr{leff} is
$S=S_W+S_A$ where
\be
S_W=\beta_G\sum_x\sum_{i<j}(1-\fr12\tr P_{ij})  \la{sw}
\ee
is the standard SU(2) Wilson action and, in continuum
normalization ($A_0=A_0^aT_a$) and for $\bar\lambda_A=0$,
\ba
S_A&=&\sum_{x,i}2 \biggl[\tr aA_0^2(x)-\tr aA_0(x)U_i(x)
A_0(x+i)U_i^\dagger(x)\biggr]\nn\\
&&+\sum_x\biggl[(m_Da)^2\tr aA_0^2(x)+a\lambda_A
(\tr aA_0^2)^2\biggr], \la{sa}
\ea
where
\be
a={4\over\beta_G g_3^2}  \la{betag}
\ee
is the lattice spacing and $m_D$ is the bare mass in the lattice
scheme.
Renormalisation is carried out so that
the physical results are the same as in the $\msbar$ scheme
with the renormalized mass parameter
\be
m_D^2(\mu)=\biggl(y+{20x-10x^2\over16\pi^2}\ln{g_3^2\over\mu}
\biggr)g_3^4.
\ee
The resulting  counterterms have been
computed in~\cite{mlaine2} with the result
\be
m_D^2=m_D^2(\mu)-{\Sigma\over4\pi a}(4g_3^2+5\lambda_A)-
{g_3^4\over16\pi^2}\biggl[(20x-10x^2)(\ln{6\over a\mu}+0.09)+
8.7+11.6x\biggr],
\ee
where $\Sigma=3.1759114$. 
There are no higher order corrections to this relation
in the limit $a\to 0$. 
For 1-loop $O(a)$-improvement,
see~\cite{moore_a}.

The action can be written in different forms by rescalings of $A_0$.
For example, replacing $A_0$ by an anti-Hermitian
matrix~\cite{reisz,reisz3}
by $aA_0^2\equiv -\beta_A \tilde{A}_0^2/4$, gives an action of the form
\be
S_A=\beta_A\sum_{x,i}\fr12
\tr \tilde{A}_0(x)U_i(x)\tilde{A}_0(x+i)U_i^\dagger(x)+
\sum_x\biggl[-\beta_2\fr12
\tr \tilde{A}_0^2+\beta_4(\fr12\tr \tilde{A}_0^2)^2\biggr],
\la{salattice}
\ee
where
\be
{\beta_2\over\beta_A}=3+{8y\over\beta_G^2}-
{\Sigma(4+5x)\over2\pi\beta_G}-{1\over2\pi^2\beta_G^2}
\biggl[(20x-10x^2)(\ln{\fr32 \beta_G}+0.09)+
8.7+11.6x\biggr]
\ee
and
\be
\beta_4={x\over\beta_G}\beta_A^2.
\ee
Since there are two dimensionless parameters, one more arbitrary choice
is possible. In fundamental Higgs theories one often scales the
coefficient
of the quadratic term to unity; here we choose $\beta_A=\beta_G$.

As the final relation, one wants to determine the continuum theory
value of $\langle A_0^aA_0^a\rangle$ when doing simulations with
the lattice action (\ref{sw})+(\ref{salattice}) with $\beta_A=\beta_G$.
The answer is
\be
{\langle A_0^aA_0^a\rangle\over 2 g_3^2}={\beta_G^2\over8}
\biggl[-\langle\fr12\tr \tilde{A}_0^2\rangle-{3\Sigma\over4\pi\beta_G}-
{3\over\beta_G^2\pi^2}\biggl(\log{3\beta_G\over2}+0.67\biggr)\biggr].
\la{condvalue}
\ee

\subsection{Simulations}

The aim of the simulations is as follows:
\begin{itemize}
\item Find the critical curve $y=y_c(x)$ and, in particular, its
endpoint $x_{\rm end}$. 
This is done by measuring the distributions of $\tr
A_0^2$ (or some other operator) at fixed $x$, finding a two-state
signal and the $V\to\infty, a\to 0$ limit of the value of $y$ at which
this happens,
\item Measure the correlator masses on both sides of the transition
line and in the cross-over region $x>x_{\rm end}$.
\end{itemize}

It is important to estimate the required lattice volume $V$ and
constant $a$ (or $\beta_G$). In the broken phase, on the tree level we
have at least two relevant mass scales, the adjoint scalar mass
$\approx gT$ from eq.~\nr{scalarm} and the vector mass $g\langle
A_0^3\rangle=2\pi T$ from eq.~\nr{vectorm}. Thus, in order to describe
accurately the broken phase, we must demand
\be
a\ll {1\over 2\pi T}\ll {1\over gT}\ll Na.
\ee
These convert to
\be
\beta_G\gg{8\over3\pi x},\qquad N\gg{\pi\sqrt3\over4}\beta_G
\sqrt{x}.
\la{latticesize}
\ee
For the symmetric phase the scale $2 \pi T$ is absent, so that the
requirement on $\beta_G$ is not quite as demanding, $\beta_G \gg 1$.

The simulations were performed with a Cray C90 at the Finnish
Center for Scientific Computing, and the total cpu-time usage
was 760 cpu-hours.

\subsection{The phase diagram}

We locate the transition line $y=y_c(x)$ by using $\tr A_0^2$ as an
order parameter.  For each lattice size, we define the {\em
pseudocritical\,} $y$ with the following commonly used methods
\cite{Herrmann}: (a) maximum of the susceptibility $\chi(A_0^2) =
\langle(\tr A_0^2 - \langle \tr A_0^2 \rangle)^2\rangle $;
(b) minimum of the 4th
order Binder cumulant $B_L(A_0^2) = 1 -
\langle(\tr A_0^2)^4\rangle/(3\langle (\tr
A_0^2)^2 \rangle^2)$; and, when the transition is strongly 1st order,
(c) the ``equal weight'' criterion for the probability distribution
$p(\tr A_0^2)$.
For any finite volume, these criteria give different pseudocritical
values of $y$, but
they all extrapolate to the same limit when
$V\rightarrow\infty$.  This extrapolation is shown in
Fig.~\ref{vextrapolation} for $x=0.20$.

\begin{figure}[tb]

\vspace*{-0.5cm}

\epsfysize=18cm
\centerline{\epsffile{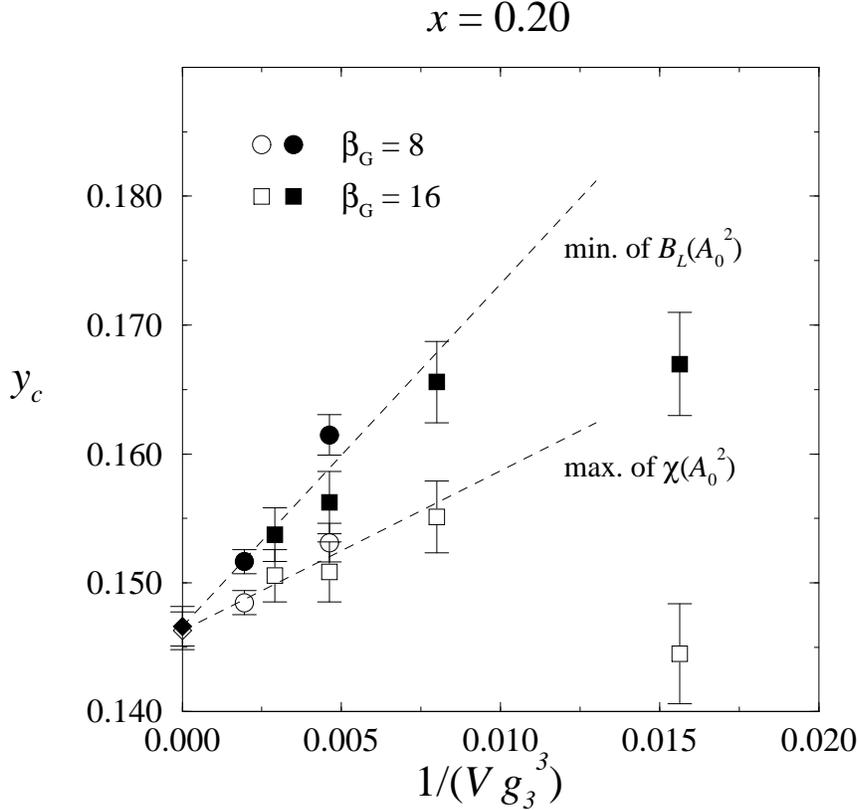}}

\vspace*{-6cm}

\caption[a]{The $y_c$ extrapolation to $V\rightarrow\infty$ for
$x=0.20$.  Within the statistical errors, there is no lattice
spacing $\beta_G$ dependence.}
\la{vextrapolation}
\end{figure}

The analysis is repeated for several $\beta_G$, corresponding to
different lattice spacings through eq.~\nr{betag}.  In our case,
as long as the conditions \nr{latticesize} are satisfied, we found
no appreciable lattice spacing dependence in the results.  This
is also evident in Fig.\ref{vextrapolation}.

Our result for the phase diagram of the continuum SU(2)+adjoint Higgs
model is shown in Fig.~\ref{ycdata}. As already pointed out in
\cite{ph}, the phase diagram consists of a first-order line which
terminates, so that the two ``phases'' are analytically connected. We
find the endpoint to be close to $x_{\rm end}=0.3$. As can be seen from the
figure, the results are independent of $\beta_G$ well within the
statistical accuracy. However, the infinite volume extrapolation is
essential, as shown by some large but finite volume points in
Fig.~\ref{ycdata}.

\begin{figure}[tb]

\vspace*{-0.5cm}

\epsfysize=18cm
\centerline{\epsffile{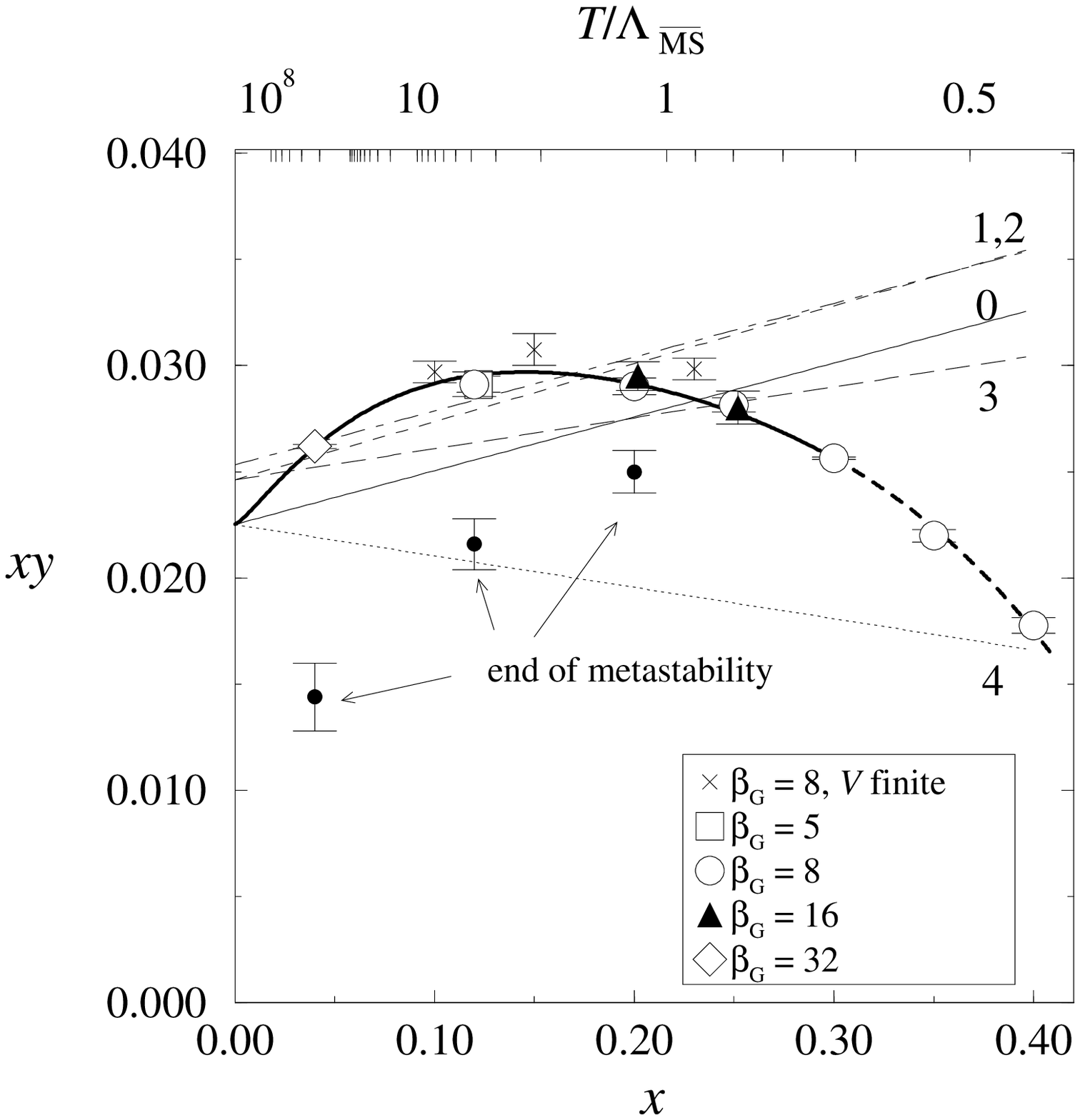}}

\vspace*{-7cm}

\caption[a]{Data points in the limit $V\to\infty,a\to0$ for
the critical  curve $y=y_c(x)$  (multiplied by $x$).  The thick line is
the 4th order fit (\ref{fit}) to the $V=\infty$ extrapolated data.  
The dashed line
marks the region where the transition turns into a cross-over.
The straight lines are the 4d$\to$3d curves of eq.~(\ref{y_dr2})
marked by the value of $N_f$. The top scale shows the values of
$T/\lambdamsbar$ corresponding to the values of $x$ for $N=2,N_f=0$,
obtained using eq.~(\ref{xT}). The physical implications of the figure
are discussed in Sec.~6.}
\la{ycdata}
\end{figure}

The thick transition line $y_c(x)$ in Fig.~\ref{ycdata} has been obtained
with a  fit to the $V\rightarrow\infty$ -extrapolated
results, as
\be
x y_c = 2/(9 \pi^2)(1 + 9/8 x + A x^{3/2} + B x^2 + C x^{5/2} + D x^3),
\label{fit}
\ee
where $A=45.1(2.9),~B=-214.7(20.2),~
C=350.7(44.5),~D=-206.7(31.4)$. The fit has been constrained to join
the  curve $x y_\dr(x)$ and its slope when $x\rightarrow 0$. We used
fractional powers of $x$ in the fit since in the perturbative expansion
of the critical curve terms like $g_3^2/m_D \sim {\sqrt{x}}$ 
appear. There can also be logarithms in the expansion.

It is interesting that the behaviour of the 3d SU(2) + adjoint Higgs
model is quite different from that of the 3d SU(2) +
fundamental Higgs theory studied earlier \cite{nonpert}. First, in the
fundamental case the value of $y_c$ was smaller than the 2-loop
perturbative result for all values of $x$, while in the adjoint case
the situation is opposite for $x < 0.2$ (compare Figs.~\ref{ycritical}, 
\ref{ycdata}). 
Second, the magnitude of the
higher order perturbative or non-perturbative effects is much larger in
the adjoint case. To get a feeling of the difference, suppose that the
deviation from the perturbative result comes entirely from a
non-perturbative energy shift at $A_0=0$ \cite{mspl,nonpert},
\be
\delta V = \frac{1}{12}A_F g_3^6,
\ee
where $A_F$ is some constant to be determined numerically. For the
fundamental case this was estimated to be $A_F \simeq -0.08$
at $x\sim 0.06$. A first
order estimate of the change in the critical value, 
\be
\fr12 \delta m_D^2 A_0^2 = - \delta V,
\ee
gives in the present case
\be
A_F \simeq - \frac{8}{3 \pi^2} \frac{\delta y_c}{x^2}.
\ee
For example, for $x=0.04$ we have $A_F \sim 10$ ($\sim 100$ times
larger than in the fundamental case!) and for $x=0.12$, $A_F \sim 0.4$.
We do not have any clear understanding of this huge difference. From
the point of view of perturbation theory, both cases are quite
similar. The only difference we see is that the adjoint theory contains
instantons (monopoles) in the broken phase, while the fundamental does
not.

In Fig.~\ref{a2histograms} we show the probability distributions
$p(A_0^2)$, measured with various values of $x$ along the transition
line $y_c(x)$.  When $x\sim 0.04$, the transition is extremely
strongly first order, as evidenced by the strong separation of the two
peaks in the distribution.  Indeed, for $x \le 0.12$ we use the
multicanonical algorithm in order to enhance the tunnelling from one
phase to another -- this is absolutely necessary in order to ensure
the correct statistical weights for the phases.  When $x$ increases,
the transition becomes weaker, and at $x=0.40$ we see no sign of a
transition any more.

Due to the strong 1st order transition, there is a substantial {\em
metastability\,} range in $y$ around the transition line at small $x$.
Within this range the thermodynamically unstable state can be stable
enough so that, for practical purposes ($\sim$ measurements), it
behaves as if it were stable.  In Fig.~\ref{ycdata} we show the lower
end of the symmetric phase metastability range for $x=0.04$, 0.12 and
0.20\@.  This has been obtained by reweighting the $y_c(x)$
-distributions (Fig.~\ref{a2histograms}) to smaller values of $y$
until the peak of the distribution corresponding to the symmetric phase
vanishes.

\begin{figure}[tb]

\vspace*{0cm}

\epsfysize=18cm
\centerline{\epsffile{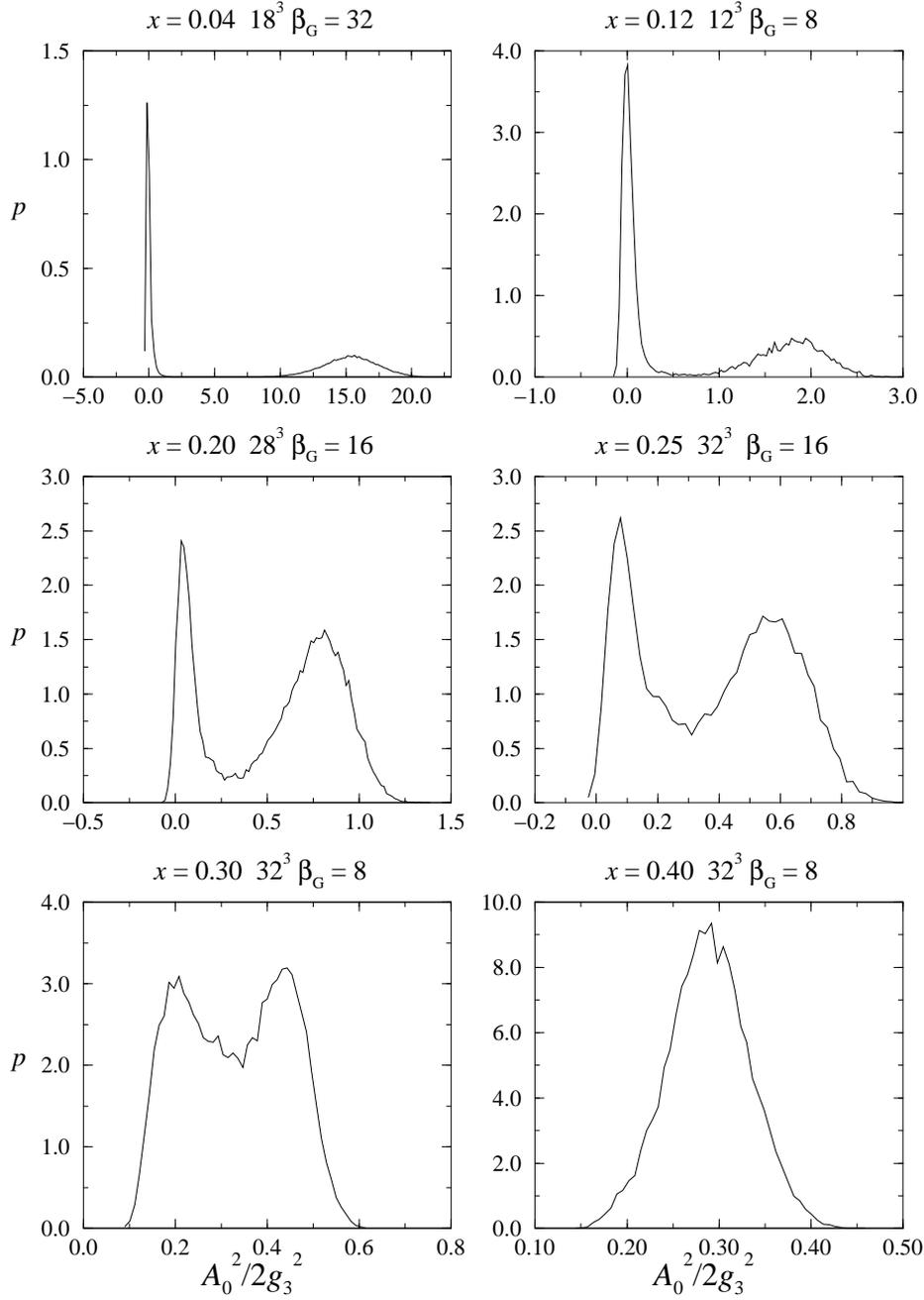}}

\vspace*{-1cm}

\caption[a]{The probability distributions of $A_0^a A_0^a/2g_3^2 =
\fr1{N_S^3}
\sum_x A_0^a A_0^a/2g_3^2$, measured at various locations on the
transition
line $y=y_c(x)$.}
\la{a2histograms}
\end{figure}

The behaviour of $\langle A_0^a A_0^a \rangle/2 g_3^2$ when crossing
the transition line or its continuation is shown in
Fig.~\ref{conddata}.

\begin{figure}[t]

\vspace*{-0.4cm}

\centerline{\epsfxsize=9cm\epsfbox{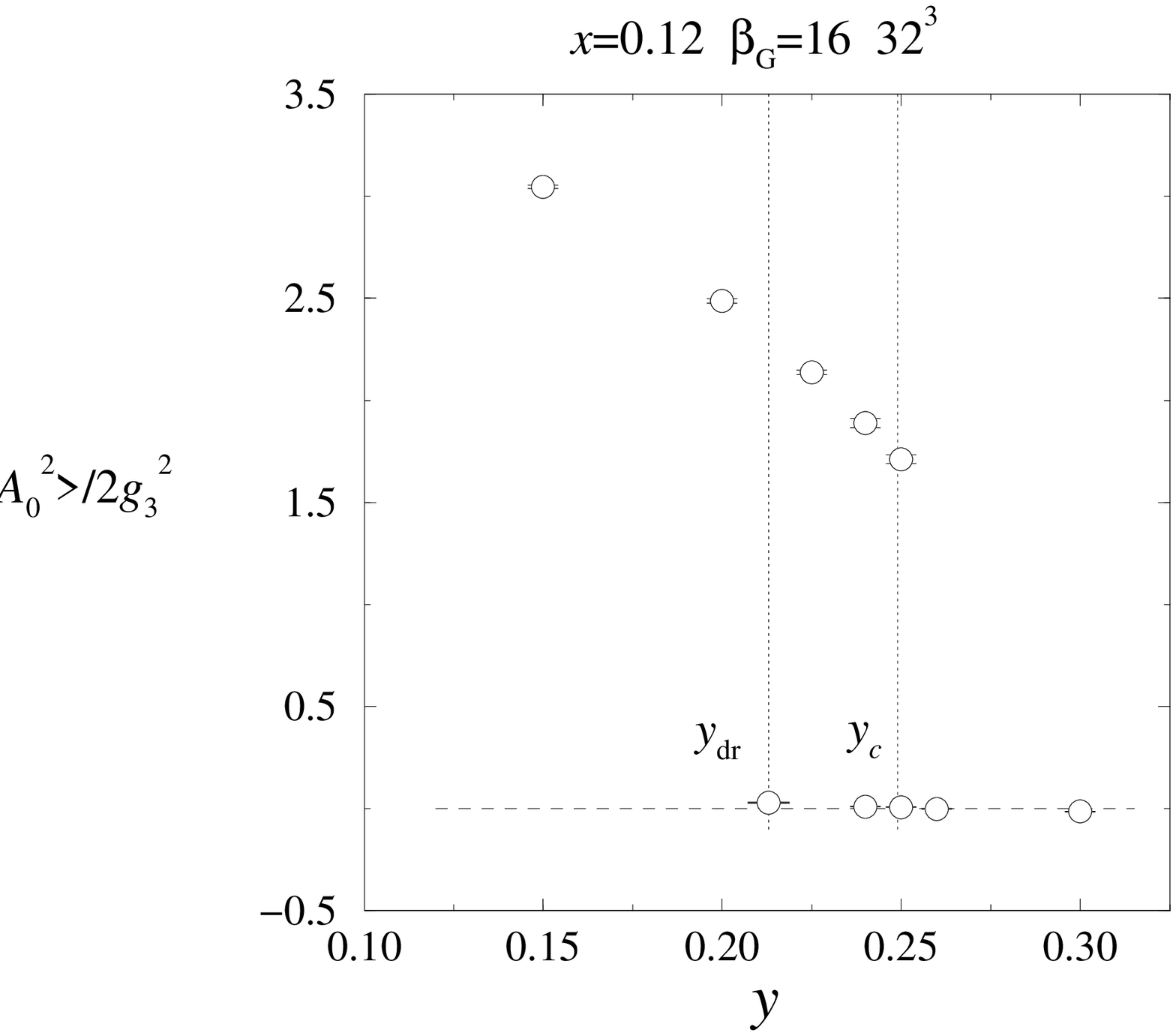}}

\vspace*{-4.4cm}

\centerline{\epsfxsize=9cm\epsfbox{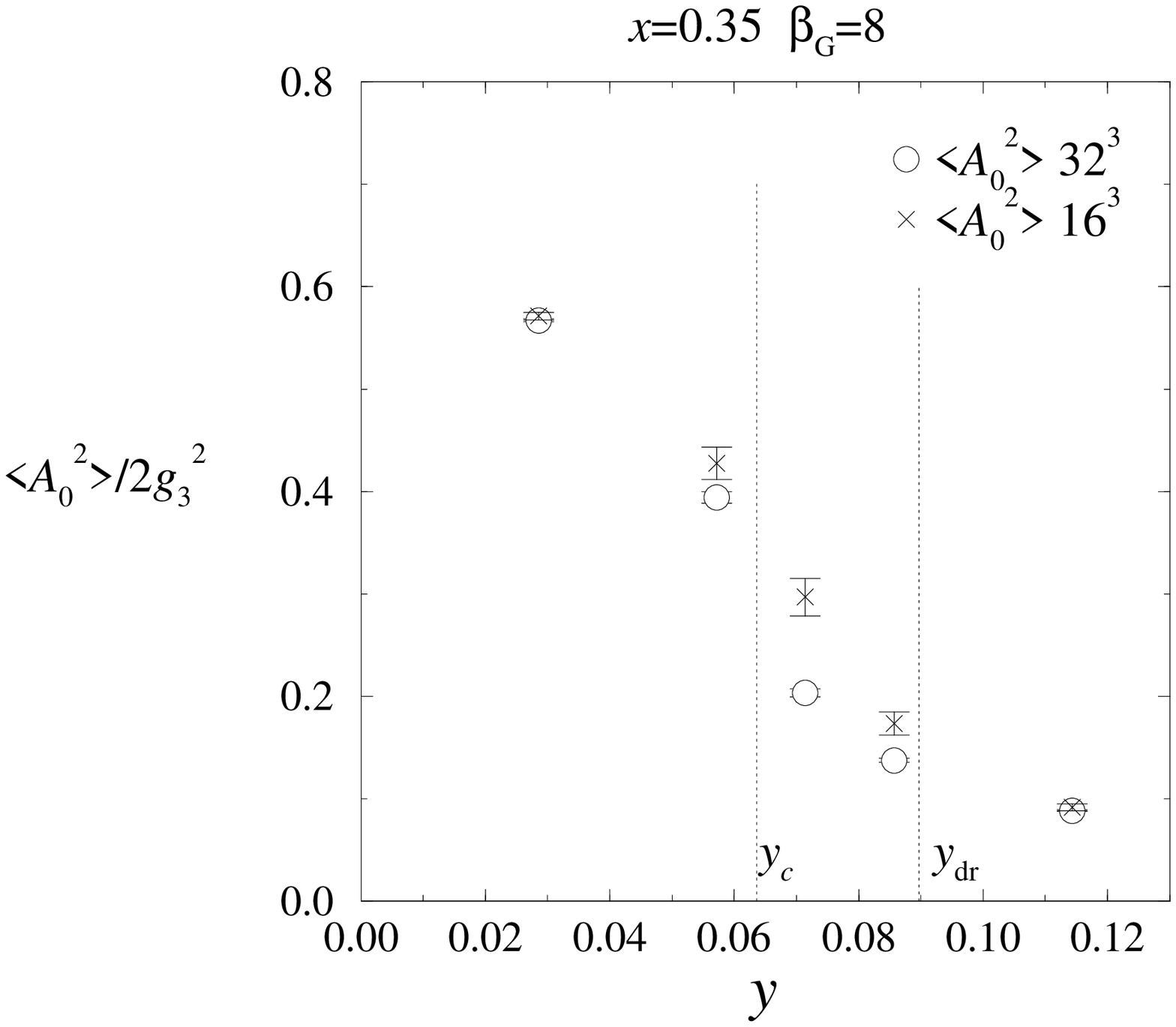}}

\vspace*{-4.4cm}

\caption[a]{The behaviour of $\langle A_0^a A_0^a\rangle/2g_3^2$ in
continuum normalization (see eq.~(\ref{condvalue})) when crossing
the phase transition at $x=0.12$ and the cross-over at $x=0.35$.
 }
\la{conddata}
\end{figure}

\subsection{Correlator masses}

The list of gauge-invariant composite operators of lowest 
dimensionality is:
\begin{itemize}
\item Dim = 1: the scalar $\tr A_0^2$,
\item Dim = 3/2:
the vector $h_i=\epsilon_{ijk} A_0^aF_{jk}^a$,
\item Dim = 2: the
scalar $(\tr A_0^2)^2$,
\item Dim = 3: the scalars $(\tr A_0^2)^3$ and
$G=F_{ij}^aF_{ij}^a$ and the tensor $F_{ij}^aF_{jk}^a$.
\end{itemize}

Perturbatively the Lagrangian \nr{leff} for $N=2$ in the symmetric
phase has 3$\times$2 massless vector ($A_i^a$) and 3 scalar ($A_0^a$)
degrees of freedom. Because of confinement, only SU(2) singlets appear
in the spectrum. Thus, the above mentioned operators represent bound
states in the symmetric phase. For example, $\tr A_0^2$ is a bound
state of two adjoint scalar ``quarks", $h_i$ is a bound state of
an adjoint quark and light glue, 
and $G$ is a scalar glueball state.

In the limit of large $y \gg 1$ the SU(2) theory under
consideration in the symmetric phase can be considered an analog of
QCD with heavy scalar adjoint quarks. Thus, a number of statements can
be made on the expected spectrum with the use 
of the heavy quark expansion.
First, the lowest energy states are
different types of glueballs with $m_G \sim g_3^2$. Next, we have a
bound state $h_i$ of the scalar quark and gluon with the mass
$m_h= m_D + O(g_3^2)$, $m_D=\sqrt{y} g_3^2$. 
The leading correction to $m_h$ can
be determined:
\be
m_h - m_D = \frac{g_3^2}{2 \pi} \log\sqrt{y} +
\frac{c}{2\pi} g_3^2 +...
\label{cconst}
\ee
The coefficient in front of the log is a result of a
perturbative computation of the pole $A_0$ mass in \cite{rebhan}, 
whereas the
constant term proportional to $g_3^2$ cannot be computed
perturbatively. In fact, the mass of this bound state coincides with
the non-perturbative definition of the Debye screening mass, discussed
in \cite{ay}. Finally, the mass of the bound state $A_0^2$ is
expected to be
\be
m(A_0^2)= 2 \sqrt{y} g_3^2 + O(g_3^2).
\la{ma02}
\ee

In the broken phase the perturbative spectrum consists of 2$\times$3
massive vectors ($W^\pm$),
$m_W \simeq 2 g_3^2/(3\pi x)$, 
2 massless vectors (the photon $\gamma$) and one scalar degree of freedom
(the Higgs) with mass $\simeq m_D$. The 2$\times$3 massive vector states
$W^\pm$ have U(1) charges and thus are not gauge invariant. Moreover,
in 3d the U(1) Coulomb potential is logarithmically divergent, and thus
the U(1) charged states (like $W$) must be absent from the spectrum.
Therefore, the following set of gauge-invariant operators should 
describe the spectrum of the theory in the ``broken" phase:
$\tr A_0^2$ for the Higgs excitation, $h_i$ for the ``photon", and $G$
for the bound state of the massive $W$'s. In fact, to obtain in the
unitary gauge $\hat A_0^a \equiv A_0^a/\sqrt{A_0^bA_0^b} =
\delta_{a3}$ a pure photon operator, one may take \cite{thooftmon}
\be
\gamma_{ij} =
\hat A_0^aF_{ij}^a-{1\over g_3}\epsilon_{abc}\hat A_0^a
(D_i\hat A_0)^b(D_j\hat A_0)^c.
\ee
However, in our measurements this operator was dominated by the first
term, yielding similar results as the
operator $h_i$. As discussed in the introduction, the broken phase
tree-level massless photon is actually replaced by a
massive pseudoscalar excitation.

In the following we constrain ourselves to the study of three different
operators, $\tr A_0^2$, $h_i$ and $G$. Our strategy is the following:
first, we are going to consider correlation functions near the 3d
transition line. This will allow us to compare masses of different
excitations in the vicinity of the phase transition, as well as to
determine correlation lengths in the symmetric phase 
(as we will discuss later, this is the 
physical phase from the 4d point of view) along the 4d $\rightarrow$ 3d
dimensional reduction curve. Second, we make measurements at some fixed
$x$ for different values of $y$ in the symmetric phase, in order to
check different hypotheses concerning the spectrum of the bound states
and in order to determine non-perturbatively the Debye screening mass and 
the $O(g^2T)$ correction to it.

The correlation functions are measured in the direction of the $x_3$
-axis, and to enhance the projection to the ground states, we
use {\em blocking\,} in the $(x_1,x_2)$-plane. The fields are
recursively mapped from blocking level $(k) \rightarrow (k+1)$, so that
the fields on the $(k+1)$-level lattice are defined only on even points
of the $(k)$-level lattice on the $(x_1,x_2)$-planes:
\be
  A_0^{(k+1)}(y) = \fr15 A_0^{(k)}(x) +
        \fr15 \sum_{i=\pm 1,2} U^{(k)}_i(x) A_0^{(k)}(x+i)
U^{(k)\dagger}_i(x)
\ee
and ($i = 1,2, j\neq i$)
\ba
  U^{(k+1)}_i(y) &=& U'^{(k)}_i(x)U'^{(k)}_i(x+i), \la{ublock1} \\
  U'^{(k)}_i(x)  &=&  \fr13 U^{(k)}_i(x)  +  \fr13 \sum_{j=\pm 1,2}
        U^{(k)}_j(x) U^{(k)}_i(x+j)
                U^{(k)\dagger}_j(x+i). \la{ublock}
\ea
The scalar and vector operators above are then calculated from the
blocked fields. The blocking is repeated up to 4 times, and the
correlation functions are measured for each blocking level separately.
For each operator and coupling constant we select the blocking level
which yields the best exponential fits; typically level 3 or 4.

Because we expect the vector operator $h_i$ to yield a very light mass
value in the broken phase, we measure it in the lowest non-zero
transverse momentum channel (for all of the blocking levels), in
addition to the zero momentum channel:
\be
 O_3(x_3) = \fr1{N_S^2} \sum_{x_1,x_2} A_0^aF_{12}^a e^{i 2\pi
x_1/N_S}.
\ee
The screening mass is then extracted from the asymptotic behaviour
of $\langle O_3(0)O_3(x_3)\rangle\! \propto \exp[-W x_3]$, where
\be
W = \sqrt{(2\pi/N_S)^2 + m_\gamma^2}.  \la{wee}
\ee

\begin{figure}[t]

\vspace*{-0.4cm}

\centerline{\epsfxsize=9cm\epsfbox{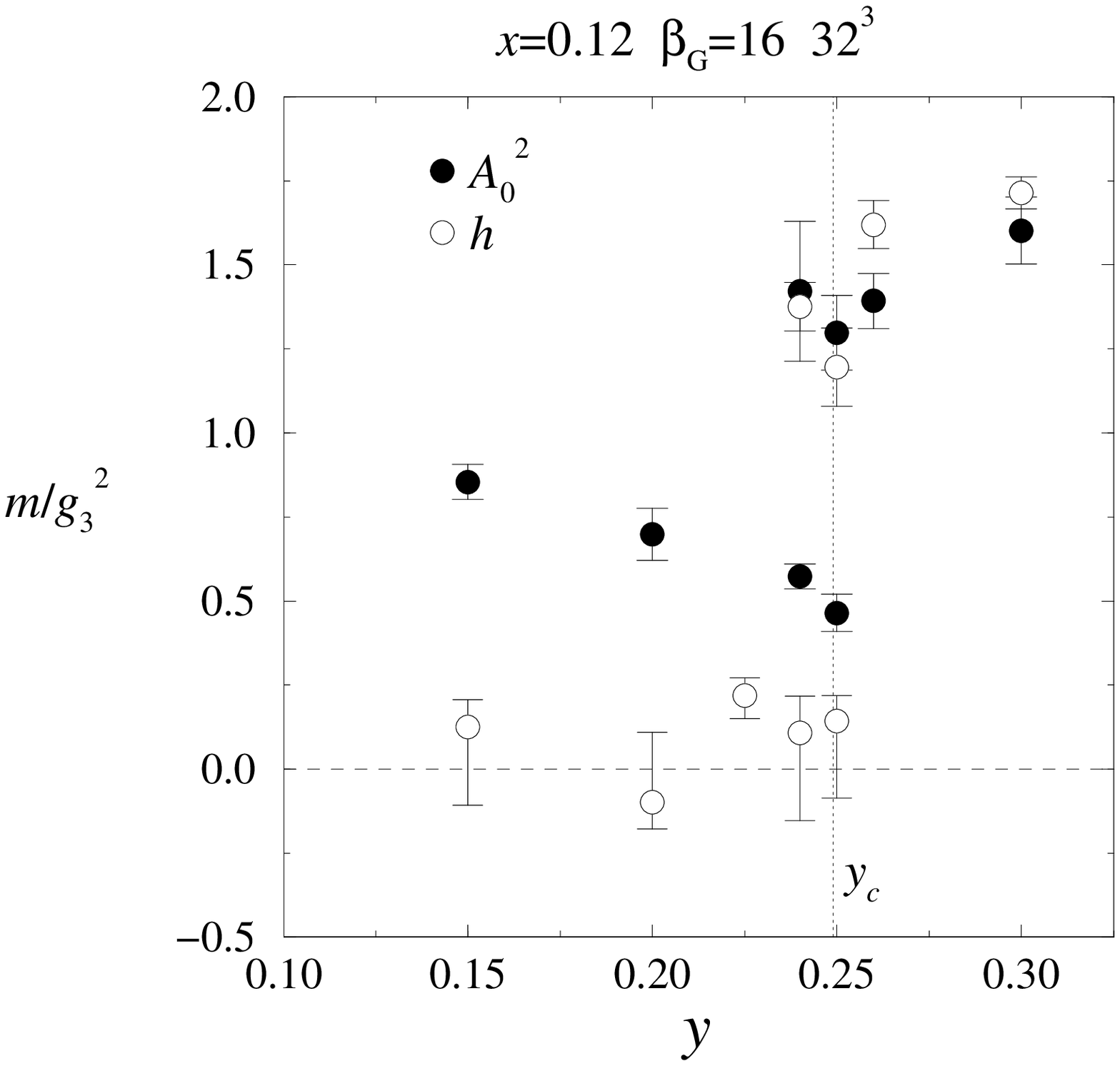}}

\vspace*{-4.4cm}

\centerline{\epsfxsize=9cm\epsfbox{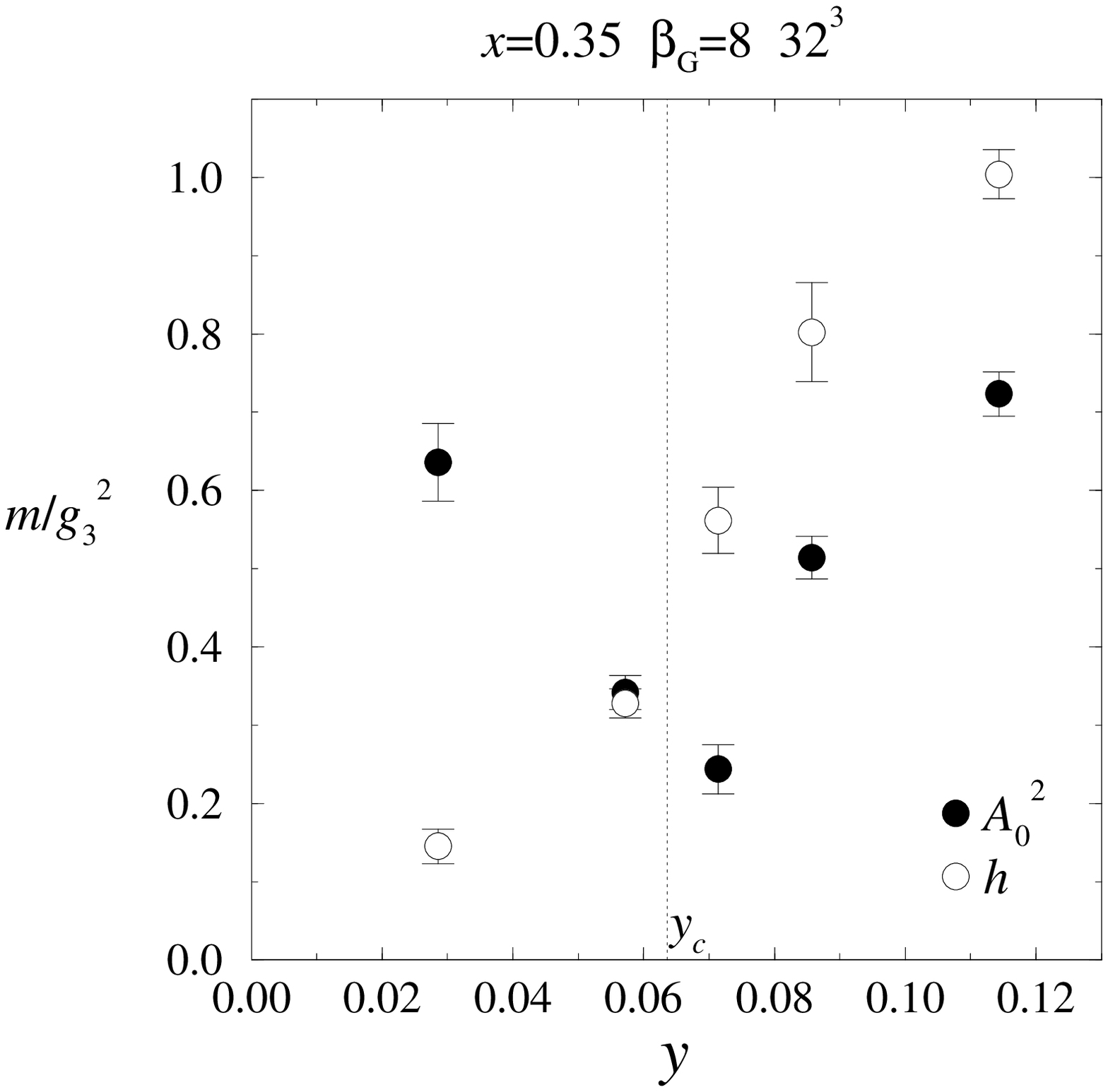}}

\vspace*{-4.4cm}

\caption[a]{The correlator masses of the operators $\tr A_0^2$ and
$h_i=\epsilon_{ijk}\tr A_0F_{jk}$ measured when crossing
the phase transition at $x=0.12$ and the cross-over at $x=0.35$.
 }
\la{massdata}
\end{figure}

The results for the masses both in the 1st order phase transition
region ($x=0.12$) and the cross-over region ($x=0.35$) are shown in
Fig.~\ref{massdata}. At the phase transition the masses exhibit a
prominent discontinuity. In the broken phase $y<y_c$ there is a photon
of very small mass and a scalar the mass of which is close to 
the mass parameter $m_D$; 
in these units $\sqrt{2/3}gT/g^2T= \sqrt{2/(9\pi^2x)}$. In the
broken phase only the non-zero transverse momentum operator gives
statistically significant results for $h_i$. 
The negative mass values in
Fig.~\ref{massdata} notify negative $m_\gamma^2$-values, as obtained
from the relation \nr{wee}. The photon mass is statistically consistent
with zero but, as discussed already in \cite{ph}, a small non-zero
photon mass is not excluded. This is required for the analytic connection
between the two ``phases''; otherwise the photon mass could act as an
order parameter separating the phases (this, in fact, is what happens 
in the U(1)+Higgs case, the Ginzburg-Landau theory \cite{kklp}).
In the symmetric phase $y>y_c$
there is a scalar $A_0-A_0$ and a vector $A_i-A_0$ bound state of
rather large mass. In the cross-over region the mass of the scalar
state dips at $y=y_c$ while that of the vector state increases
monotonically.

\begin{figure}[t]

\vspace*{-1.0cm}

\hspace{1cm}
\epsfysize=18cm
\centerline{\epsffile{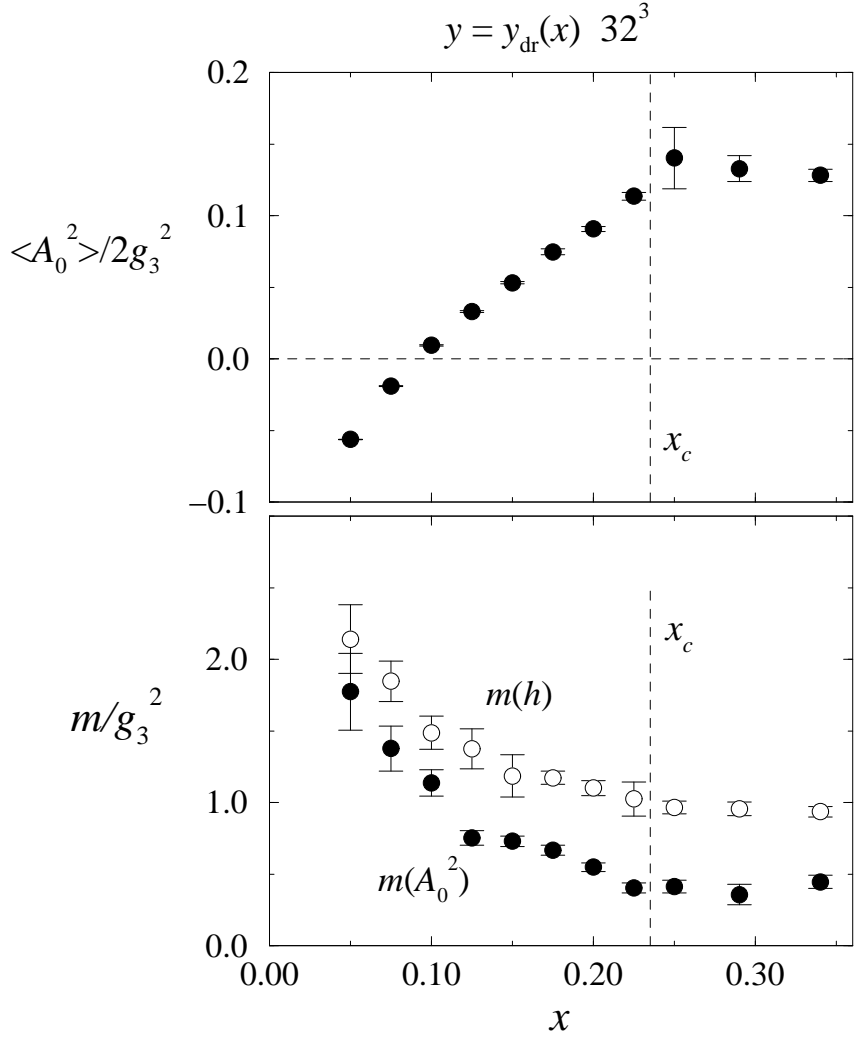}}

\vspace*{-4cm}

\caption[a]{
The order parameter
$\langle A_0^aA_0^a\rangle/2 g_3^2$
and the scalar and vector masses $m(A_0^2)$, $m(h)$ in the
symmetric phase along $y=y_\dr(x)$.  When $x < x_c \approx 0.235$, the 
symmetric phase is metastable.  The $x<x_c$ -points have been calculated
with $\beta_G=8$, $x>x_c$ with $\beta_G=12$.}
\la{meta}
\end{figure}

Now, consider the masses of $A_0^2$ and $h_i$ in the symmetric phase
along the dimensional reduction curve $y=y_\dr(x)$, Fig.~\ref{meta}.
These measurements are possible since for $x<0.235$ the symmetric phase
is metastable with a sufficiently long tunneling time, while at
$x>0.235$ it is absolutely stable. As one can see the expectation
$m(A_0^2)/m(h) \simeq 2$ is strongly violated in the whole region of
$x$ studied, $x > 0.05$. This indicates that the corrections to the
leading order results \nr{cconst}, \nr{ma02}
are large. In order to clarify this point, we
study different correlators at large values of $y$ and and fixed x,
where the heavy quark expansion is expected to be valid.

In Fig.~\ref{01} 
we show the dependence of the masses
of the bound states on $y$ at different values of $x$,
$x=0.01, 0.04, 0.10$. As expected, the glueball mass does practically 
not depend on $y$ or $x$, while the $A_0^2$ and $h_i$ masses have some
dependence on $x$. Moreover, at large $y$, $m(A_0^2)> m(h)$ as it
should be. 
A fit to the data allows to fix the unknown coefficient $c$ in
(\ref{cconst}). Only sufficiently large values of $y$ should be used,
$\sqrt{y} > c$, to make sure that the correction is smaller than the
leading contribution. We found for $y > 2$:
\ba
x &=& 0.10 : c = 1.74(16)\cdot(2 \pi),\nonumber \\
x &=& 0.04 : c = 1.53(17)\cdot(2 \pi),\nonumber \\
x &=& 0.01 : c = 1.39(20)\cdot(2 \pi).
\ea
This indicates a slight $x$-dependence of the parameter $c$, which
is perhaps a result of higher order corrections.
A linear in $x$ fit works very well, and the 
intercept represents a realistic estimate for c,
\be
c= 1.36(18)\cdot(2 \pi).
\ee
This is basically the same as the result of the x=0.01 point.

The large value of the coefficient $c$ explains why the heavy quark
expansion does not work for the values of $x$ considered in
Fig.~\ref{meta} along the dimensional reduction curve $y=y_\dr(x)$.
The adjoint ``quark" can be considered heavy only if $\sqrt{y}>
{c}/(2\pi)$, which gives $x < {8}/(9 c^2)\simeq 0.01$.

\begin{figure}

\vspace*{-2.5cm}

\hspace{1cm}
\epsfysize=12cm
\centerline{\epsffile{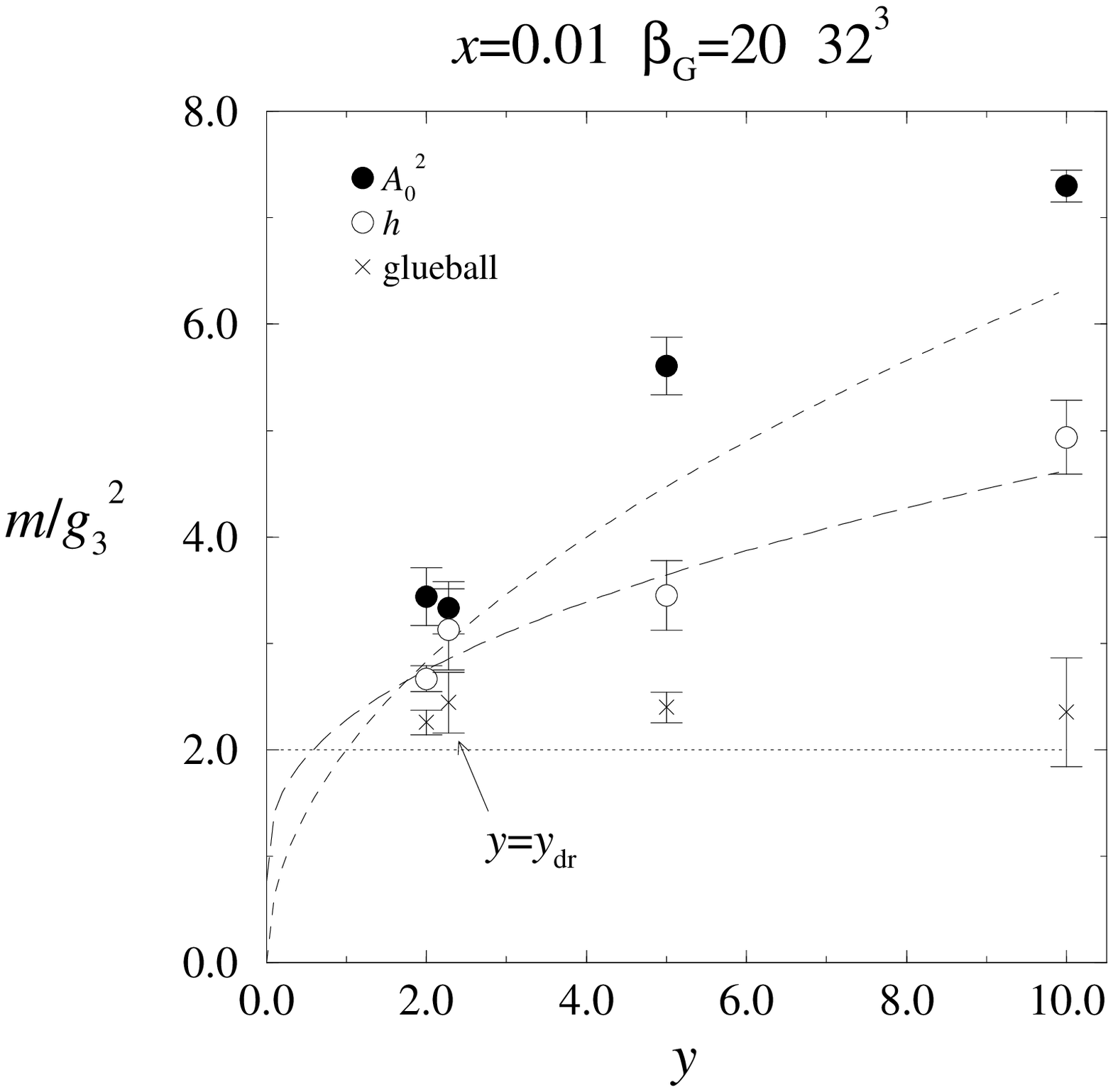}}

\vspace*{-4.6cm}

\epsfysize=12cm

\centerline{\epsffile{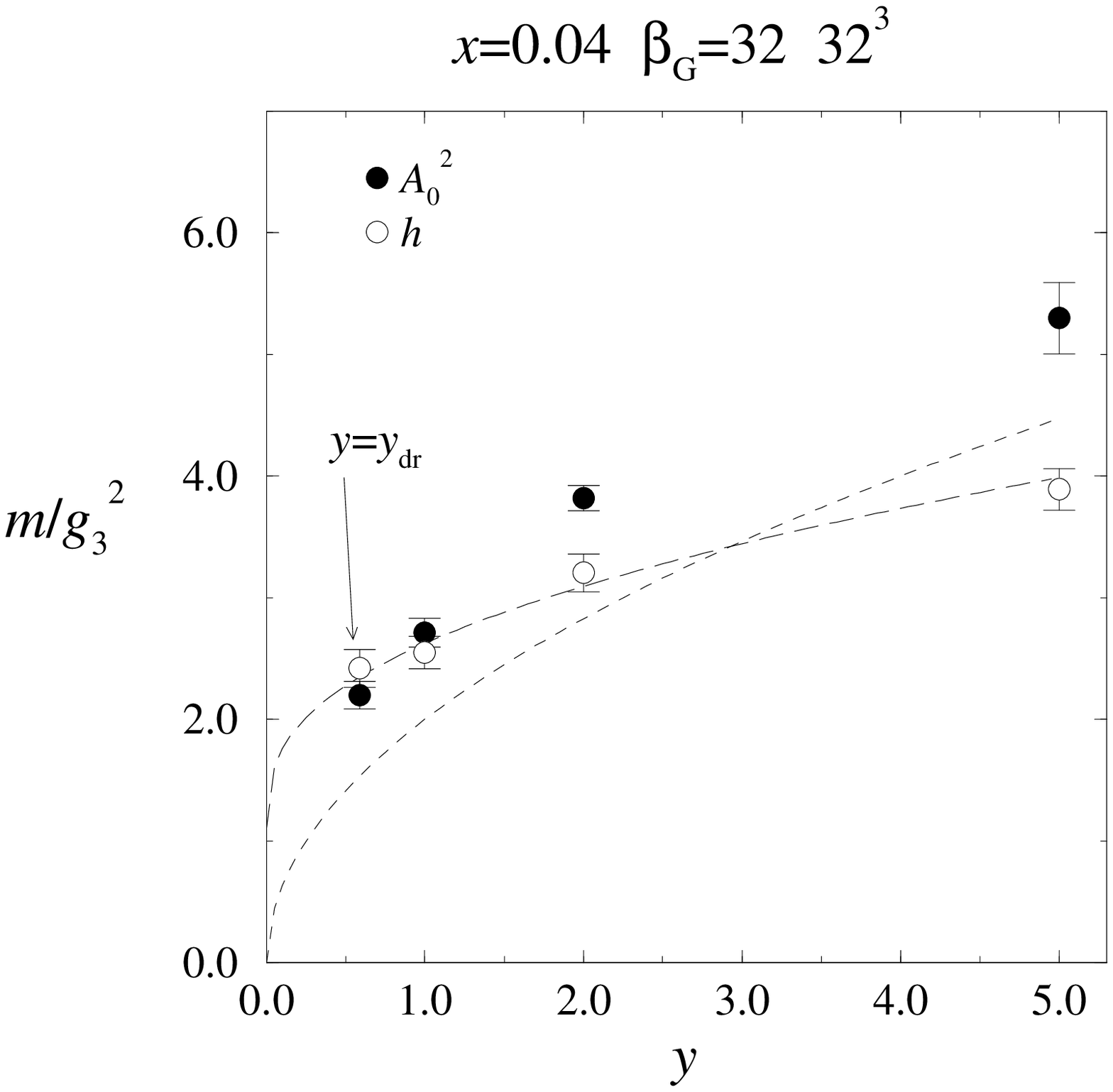}}

\vspace*{-4.6cm}

\epsfysize=12cm

\centerline{\epsffile{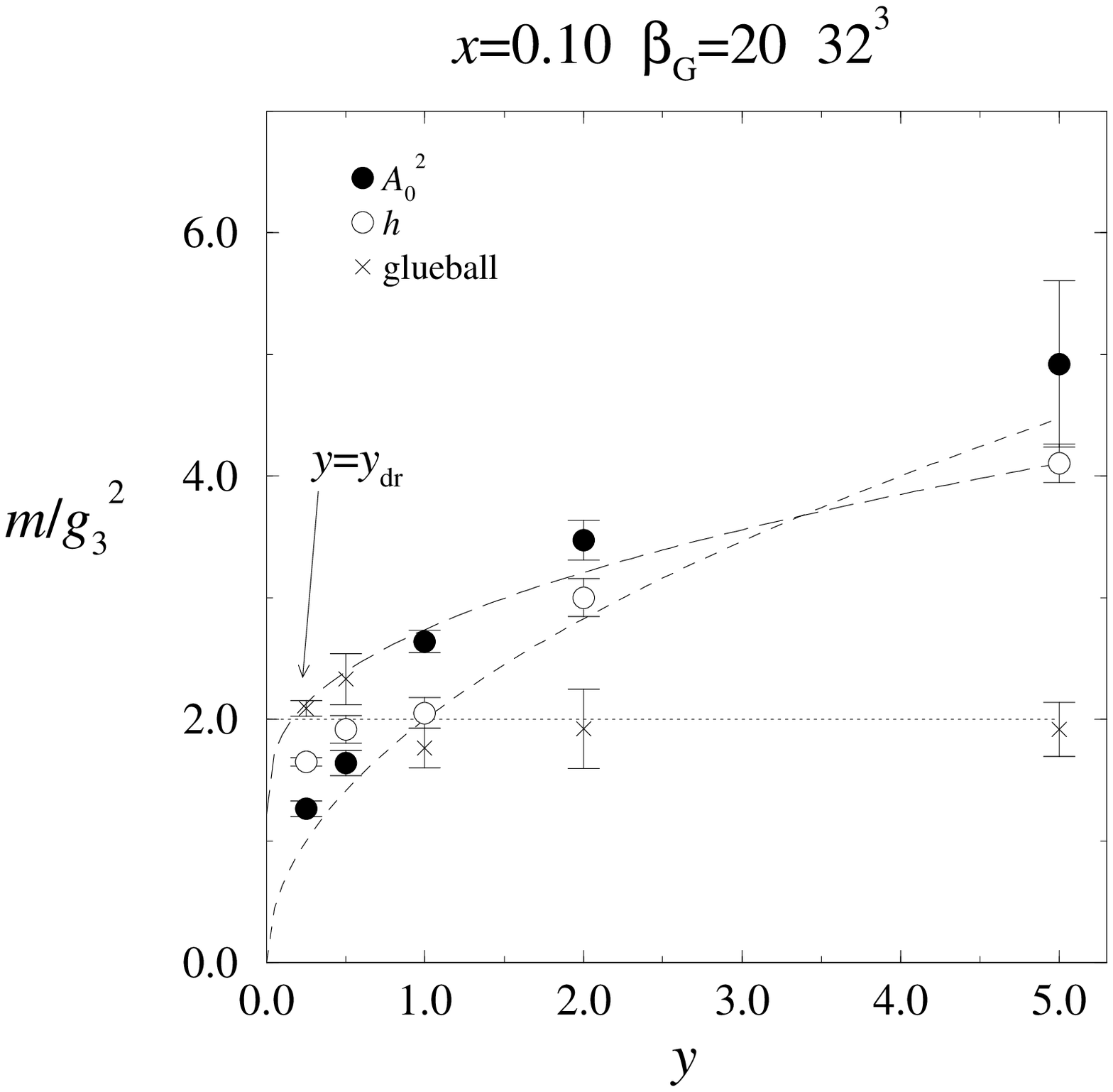}}

\vspace*{-4.5cm}

\caption[a]{Scalar, vector and glueball bound state masses for
$x=0.01,~0.04,~0.10$.}
\la{01}
\end{figure}

\section{The physical phase of the 3d theory}

The previous simulation results were entirely for the 3d effective
SU(2)+adjoint Higgs theory as such. We can now return to the main
question: what is their relevance for the 4d finite temperature $N=2$
QCD?

According to the 2-loop dimensional reduction in Sec.~2, the physical
4d theories, defined by the values of $T/\lambdamsbar,N=2,N_f$, lie on
the straight lines $y_\dr(x)$ (eq.(\ref{y_dr2})) plotted in
Fig.~\ref{ycdata}. Consider first the quarkless case, $N_f=0$. One
observes that for almost all the range 
$x<x_{\rm end}\sim 0.30$, one has
$y_\dr(x)<y_c(x)$ as already suggested by the 2-loop computations in
Fig.~\ref{ycritical}. Thus superficially, the finite
temperature 4d theory corresponds to the broken phase of the 3d theory.
This was the conclusion reached in \cite{polonyi} (the status of which
is reviewed in \cite{A0review}).

However, this conclusion cannot be right. The 3d broken phase is a
remnant of the 4d Z(2)-symmetry as discussed in Sec.~4, and the
derivation of the effective theory is not reliable there. Instead, the
physical phase must be the (in the 4d language, equivalent) symmetric
phase where the 3d theory can be trusted. Indeed, at $y=y_\dr(x)$ the
symmetric phase of the 3d theory is metastable in the whole region of
$x$, and all the observables can be measured within the metastable
phase, see Fig.~\ref{meta}. That the symmetric phase is the physical
phase of the 3d theory was the conclusion reached in \cite{reisz3}, as
well; however, there the 3d symmetric phase was seen to be stable. The
difference is due to increased accuracy in relating continuum and
lattice results. If one would be able to construct a 3d theory which
fully respects the Z($N$) symmetry, then the metastable
phase should be promoted to a stable phase in coexistence with its Z(2)
companion.

\begin{figure}[t]

\vspace*{-1.0cm}

\hspace{1cm}
\epsfysize=18cm
\centerline{\epsffile{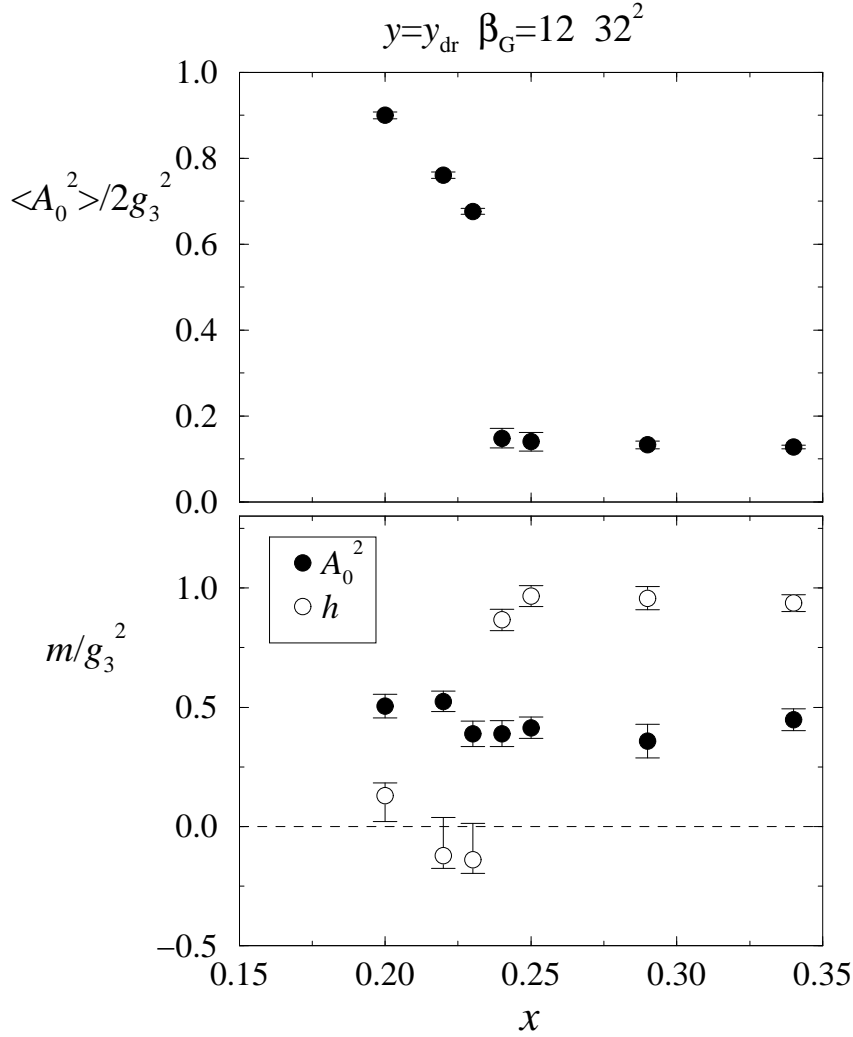}}

\vspace*{-4cm}

\caption[a]{
The order parameter $\langle A_0^a A_0^a\rangle/2g_3^2$ and the
scalar and vector masses $m(A_0^2)$, $m(h)$ in the
{\it stable} phase along $y=y_\dr(x)$. For $x<0.235$ the stable phase
is the broken phase, for $x>0.235$ the symmetric phase 
(see Fig.~\ref{ycdata}).
As discussed in Sec.~6, at $x<0.235$ it is rather the metastable
symmetric phase which is physical.}
\la{ydr_yc}
\end{figure}

What happens when one goes to smaller $T$ (larger $x$), towards $T_c$?
Then the 3d theory is becoming less reliable even in the symmetric
phase, for several reasons. First, the perturbative expansion for the
parameters of the 3d theory is becoming less reliable, see
eq.~\nr{ydrntwo}.
Moreover, the higher order operators are becoming more important since
the mass hierarchy between the scales $g_3^2, \pi T$ is lost due to
increasing $g^2(T)$ at $T/\lambdamsbar < {\rm \mbox{a few}}$, see
Sec.~2. Indeed, the masses of the excitations in the 3d theory are
$\sim g_3^2$, see Fig.~\ref{meta}, so they need then no longer be the
dominant infrared excitations.

In Fig.~\ref{ydr_yc}, the order parameter and the masses have been
plotted in the {\it stable} phase corresponding to $y=y_\dr(x)$. One
sees that at $x\sim 0.235$ the symmetric phase becomes stable. But this
corresponds to $T/\lambdamsbar\approx0.9$, where the 3d theory is no
longer reliable and the QCD phase transition has already taken place.
Thus this ``transition'' is not related to the 4d QCD transition.

Consider then the endpoint value, $x_{\rm end}\sim 0.30$ (see
Fig.~\ref{ycdata}). This corresponds to $T/\lambdamsbar\approx 0.6$ and
is again below the value $T_c/\lambdamsbar=1.23(11)$ determined
\cite{fhk} for the SU(2) phase transition with 4d simulations. Yet the
difference is not that large, and it is tempting to conjecture that if
one had a 3d theory fully respecting the Z($N$) symmetry, then the
corresponding endpoint might represent the QCD phase transition.

Consider finally the $N_f$ dependence. 
Fermions do not appear as
dynamical fields in the effective theory, and their only effect is
through the parametric dependence on $N_f$.
At very small $x$ (very large
$T$) the curves in Fig.~\ref{ycdata} imply that the system lies in the
symmetric phase near $A_0=0$. This result is in agreement with the fact
that quarks break the Z(2) symmetry: the second minimum becomes
metastable for $N_f=1$ and disappears for $N_f\ge3$. 
At very large $N_f$, on the other hand, 
the perturbative expansion for the curves $y=y_\dr(x)$ gets worse.

\section{Discussion}

Dimensional reduction allows to construct an effective field theory for
high temperature QCD, which is reliable 
as long as $g_3^2\ll \pi T$, implying
$T\gsim {\rm \mbox{a
few}}\times T_c$.  However, 
the effective super-renor\-mali\-zable 3d gauge+adjoint
Higgs theory does not describe the confine\-ment-deconfine\-ment phase
transition.

The usual ``power counting" picture of correlation lengths in high
temperature QCD says that the longest scale is related to the magnetic
sector of the theory and is of order $(g^2T)^{-1}$. The shorter
scale, $\sim (g T)^{-1}$, is associated with Debye screening. Our study
shows that this picture is correct only at extremely high
temperatures. First, the purely magnetic scale comes from the 
lowest glueball mass
in a pure SU(2) theory, $m_G \sim 2 g_3^2$. This should be
compared with the leading order
Debye mass $m_D^2 \sim \fr23 g^2 T^2$. The requirement
$m_D>m_G$ tells that only at $x < 0.005$ are the ``magnetic" effects 
numerically smaller than the ``electric" ones. A similar estimate follows
from the requirement that the non-perturbative corrections to the Debye
screening mass are on the level of, say, $50$\,\% of the tree-level value.
In terms of temperature, this gives ridiculously large numbers. For
example, for pure SU(2) this means $T > 10^{35}\lambdamsbar$.

Thus, in the physically most interesting region somewhere above the
critical temperature, the ``naive" picture is wrong. The
longest correlation length corresponds to the $0^{++}$ bound
state $A_0-A_0$ (in 3d language). In terms of 4d variables, this correlation
length can be found from the analysis of Polyakov lines,
\be
\langle L({\bf x})L^\dagger(0)\rangle \sim
\exp[-m(A_0^2)|{\bf x}|]/|{\bf x}|,
\ee
where
\be
L({\bf x}) = \tr \Omega({\bf x}),\quad
\Omega({\bf x}) = {\cal P}\exp[ig\int_0^\beta
\! d\tau\, A_0(\tau,{\bf x})].
\ee

The second largest correlation length is associated with Debye screening. 
In 3d language, it is related to the operator $h_i$.  The corresponding 4d
operators may be found in \cite{ay}. The non-perturbative Debye
screening mass is about a factor $3$ larger than the lowest order
estimate up to temperatures $T \sim 200\lambdamsbar$. The fact that
the Debye mass is non-perturbative does not allow a reliable integration
out of the $A_0$ field for the construction of the simplest effective
field theory, containing only the scale $g_3^2$, unless the temperature of
the system is extremely large.

Finally, the pure static glue sector has an even larger mass scale.
In pure SU(2) and SU(2)$ + $fundamental Higgs 
theories, $m_G\simeq 1.6 g_3^2$~\cite{phtw}; 
in the present theory we have measured $m_G$ 
to be almost the same.

It remains to be seen whether this modification of the standard picture
of the high-temperature gauge theories has applications in the
cosmological discussion of the quark-hadron phase transition or in
heavy ion collisions.

\section*{Acknowledgements}

We thank CSC-Tieteellinen laskenta Oy ---
the Finnish Center for Scientific Computing --- for computational
facilities, and C. Korthals Altes
for useful discussions.

\appendix
\renewcommand{\thesection}{Appendix~\Alph{section}}
\renewcommand{\theequation}{\Alph{section}.\arabic{equation}}

\section{}

Some useful group theoretical relations for the SU(N) generators
$T^a_{ij}$ in the fundamental representation are:
\ba
T_aT_b&=&\fr1{2N}\delta_{ab}+\fr12d_{abs}T_s+\fr{i}2f_{abs}T_s,\nn\\
T^a_{ij}T^a_{kl}&=&\fr12(\delta_{il}\delta_{jk}-\fr1N\delta_{ij}
\delta_{kl}),\nn\\
\tr T_aT_b &=& \fr12\delta_{ab}, \nn\\
\tr T_aT_bT_c&=&\fr14(d_{abc}+if_{abc}),\nn\\
\tr T_aT_bT_cT_d&=&\fr1{4N}\delta_{ab}\delta_{cd}+\fr18
[d_{abs}d_{cds}-f_{abs}f_{cds}+i(d_{abs}f_{cds}+f_{abs}d_{cds})].
\la{trtttt}
\ea
Defining the adjoint representation by $F^a_{bc}=-if_{abc}$,
one correspondingly gets
\ba
\tr F_aF_b&=&-f_{paq}f_{qbp}=N\delta_{ab},\nn\\
\tr F_aF_bF_c&=&-if_{paq}f_{qbr}f_{rcp}=i\fr{N}2f_{abc},\nn\\
\tr F_aF_bF_cF_d&=&f_{paq}f_{qbr}f_{rcs}f_{sdp}\nn\\
&=&\delta_{ab}\delta_{cd}+\delta_{ad}\delta_{bc}+
\fr{N}4(d_{abs}d_{cds}-d_{acs}d_{bds}+d_{ads}d_{bcs})
\nn\\
&=&\fr12(2\delta_{ab}\delta_{cd}+\delta_{ac}\delta_{bd}+
\delta_{ad}\delta_{bc})+\fr{N}4(d_{abs}d_{cds}-f_{abs}f_{cds}).
\ea
Eq.~\nr{trtttt} implies that for an adjoint vector $A_a,A\equiv
A_aT_a$,
\be
\fr18d_{sab}A_aA_bd_{scd}A_cA_d=\tr A^4-\fr1N(\tr A^2)^2,
\la{a*a}
\ee
so that the matrix combination on the RHS isolates the 4-point
coupling proportional to $d_{sab}d_{scd}$. For both SU(2) and SU(3),
\be
\tr A^4=\fr12(\tr A^2)^2. \la{su2su3}
\ee

\end{document}